\documentstyle[aps,prc,epsf]{revtex}
\input psfig.sty
\def\a{\alpha}
\def\b{\alpha}

\def\htm{{\hbar^2 \over m}}

\def\x{{\bf x}}
\def\y{{\bf y}}

\def\k{{\kappa}}
\def\hypfi{\theta}
\def\be{\begin{equation}}
\def\ee{\end{equation}}
\def\bea{\begin{eqnarray}}
\def\eea{\end{eqnarray}}
\def\ra{\rightarrow}
\def\ii{{\rm i}}
\def\sai{{\cal Y}_{\a,T,\tauv}(i)}

\def\saiu{{\cal Y}_{\a,1,{3\over2}}(i)}
\def\said{{\cal Y}_{\a,0,{1\over2}}(i)}
\def\sait{{\cal Y}_{\a,1,{1\over2}}(i)}
\def\sbi{\widetilde {\cal Y}_{\a}(i) }
\def\sbj{\widetilde {\cal Y}_{\a}(j) }
\def\sbk{\widetilde {\cal Y}_{\a}(k) }

\def\d{d}
\def\db{{\bf d}}

\def\ket{\rangle}
\def\tauv{{\cal T}}
\begin{document}

\title{The Kohn Variational Principle for Elastic Proton--Deuteron
       Scattering above Deuteron Breakup Threshold}

\author{  M. Viviani$^1$, A.~Kievsky$^1$ and S.~Rosati$^{1,2}$}
\address{ $^1$Istituto Nazionale di Fisica Nucleare, Via Buonarroti 2,
          56100 Pisa, Italy }
\address{ $^2$Dipartimento di Fisica, Universita' di Pisa, Via Buonarroti 2,
          56100 Pisa, Italy }

\maketitle

\begin{abstract}
The Kohn variational principle is formulated for calculating
elastic proton--deuteron scattering amplitudes at
energies above the deuteron breakup threshold.
The use of such a principle with an
expansion of the wave function on the
(correlated) hyperspherical harmonic basis
is discussed.
\end{abstract}


\bigskip

\noindent{key words: N-d scattering, Kohn variational principle, 
          S-matrix, breakup}

\bigskip

\section{Introduction}
\label{sec:intro}

In recent years, a series of accurate calculations of $N-d$
scattering observables below the deuteron breakup threshold (DBT)
by using the Kohn variational principle (KVP) have been
reported~\cite{KRV94,KRTV96}. The resulting elastic $n-d$ phase
shifts were found to be in strict agreement with those determined
via accurate solutions of the Faddeev equations
(FE)~\cite{benchII}. The interest in this reaction has been revived
consequently on the observation of large disagreements between
theoretical predictions and experimental data for the
$p-d$ and $n-d$ $A_y$ and $iT_{11}$ polarization
observables~\cite{report,KRTV96}. Since the 
solution of the $N-d$ Schroedinger equation can be obtained with
high precision, the disagreements can be traced back to the
current models for the nucleon--nucleon (NN) and three--nucleon (3N)
interactions. Therefore, the detailed studies of this process
can be used to get information on the microscopic nuclear
interaction. 

Above the DBT, the FE have been accurately solved  for the $n-d$
reaction (see, for example, Ref.~\cite{report} and references
therein). For reactions where there are more than one charged
particles in the final state, the proper inclusion of the Coulomb
interaction is a difficult problem. On the other hand, the proton
induced breakup reaction is more easily studied experimentally, and
in fact, there is a large amount of high precision $p-d$
scattering data  available. It should be noticed that the effect of
the Coulomb potential can be rather sizeable for various polarization
observables, in particular at small scattering angles. It is
therefore of relevance to obtain accurate theoretical estimates of
the observables of this reaction.

At present,  different approaches are applied to 
calculating the $p-d$ elastic observables above the DBT. The
method used in Refs.~\cite{ASZ78,AR94,AS96} consists of separating
the long--range from the short--range Coulomb effects. Then, the FE
in momentum space are solved by taking into account only the
short--range part of the Coulomb potential, whilst the long--range
contribution is included by ``renormalizing'' the solutions found.
This method has been pursued for  applications in both atomic
systems and $p-d$ scattering at $E=10$ MeV~\cite{AKM98,AMS98}. In
other approaches, the FE are solved using the
Coulomb-Sturmian-space representation~\cite{P99}, or by directly
solving the FE in configuration space~\cite{C99}.

The KVP was discussed in Refs.~\cite{BM72,LRS72,M74} to treat breakup
reactions without considering the case of two or more charged particles
in the final state.
In the past it has been used mainly for studying atomic
processes  (in particular,  the scattering of electrons or positrons
by hydrogen  atoms). Above the ionization threshold the direct
application of the KVP has however encountered various
problems  related to the infinite number of two--body
bound states between particles of opposite charge  and to the
complexity of the boundary conditions to be
satisfied~\cite{DN71,W74}. Various  alternative formulations 
have then been studied and the associated difficulties
analyzed~\cite{MN69,C78,A98}.

In this paper we reconsider the direct formulation of the KVP for
calculating the {\it elastic} part of the S--matrix of a $p-d$ collision
above the DBT. In such a process there is no  problem relating to
two--body excited bound states. The long--range Coulomb interaction
between the two protons introduces a distortion in the description of
the asymptotic three-nucleon outgoing wave. We have proved the
validity of such a principle,  considering trial wave functions which
satisfy ``simplified'' boundary  conditions allowing for practical
applications, even when using realistic nuclear interactions. The
results for several elastic observables  have been already reported in
refs.~\cite{KRV97,KRV99}.

The paper is organized as follows. Some definitions used later in the
paper are given in Sect.~\ref{sec:def}. In
Sect.~\ref{sec:asy}, a detailed description of the asymptotic behavior
of the $p-d$ scattering wave function (w.f.) is reported, while the
proof of the KVP is given in Sect.~\ref{sec:kvp}. The use of such a
principle in conjunction with the pair--correlated hyperspherical
harmonic expansion technique is discussed in Sect.~\ref{sec:phh},
while the conclusions are given in the last section.

\section{Definitions}
\label{sec:def}

Let us introduce the Jacobi vectors for three equal mass particles,
given by
\be
     {\bf x}_i= {\bf r}_j- {\bf r}_k\ ,\quad
      {\bf y}_i= \sqrt{4\over 3}   \left( {\bf r}_i-
      {{\bf r}_j+ {\bf r}_k\over 2}\right)\ ,
 \label{eq:jcoo}
\ee
where $i$, $j$, $k$ is a cyclic permutation of $1$, $2$ and $3$, and
${\bf r}_i$ denotes the position of
particle $i$. The distance between the pair $jk$ and the particle
$i$ is
\be\label{eq:dk}
   \db_i= \k \; {\bf y}_i\ ,\qquad  \k=\sqrt{3\over 4}\ .
\ee

For the $p-d$ system, the non--relativistic total energy operator is
written as
\be\label{eq:H}
  H = -{\hbar^2 \over m}\sum_{i=1}^{3} \nabla_i^2 +
   \sum_{j<k} V^{\rm NN}(\x_i)+W^{3{\rm N}}+V_C\ , 
\ee
with $V^{\rm NN}(\x_i)$ being the pair potential between the
particles $j$, $k$ and $W^{3{\rm N}}$ a 3N interaction term. Both
$V^{\rm NN}$ and $W^{3{\rm N}}$ are short--ranged. In particular,
$V^{\rm NN}(\x_i)$ and $W^{3{\rm N}}$ become vanishingly small when
$x_i>a_N$, where $a_N$  is a  distance somewhat larger than the
range of the nuclear interaction. Moreover, $V_C$ is the Coulomb
interaction operator (we disregard here other electromagnetic
interaction terms), namely
\be \label{eq:VC}
  V_C=\sum_{j<k} {e^2\over x_{i}} {1+\tau_{j,z} \over 2}
  {1+\tau_{k,z} \over 2} \ ,
\ee
where $\tau_{m,z}/2$ is the $z$--component of the isospin
operator of particle $m$.

For a three--body system, the hyperradius is defined as
\be
   \rho=\sqrt{x_i^2+y_i^2} \ ,
\label{eq:rho2}
\ee
and is independent of the permutation $i=1,2,3$. The hyperangular
variables are
\be\label{eq:omega}
  \Omega_i\equiv({\hat x}_i, {\hat y}_i, \hypfi_i)\ ,
\ee
where the hyperangle $\hypfi_i$ is defined by the relation
\be
    \cos \hypfi_i = x_i/\rho\ .
\label{eq:zz}
\ee

To describe a three--nucleon state with total angular momentum $J,
J_z$, the following functions are introduced
\be\label{eq:ang}
    \Bigl \{    \bigl[ Y_{\ell_\a}(\hat x_i)
        Y_{L_a}(\hat y_i) \bigr  ]_{\Lambda_\a}
       \bigl [ (s^j s^k)_{S_\alpha} s^i \bigr ]_{\Sigma_\alpha}
       \Bigr \}_{J J_z} \;
       \bigl [ (t^j t^k)_T t^i \bigr ]_{\tauv\tauv_z}
       \equiv     \sai \ ,
\ee
where the variables $s$ ($t$) specify spin (isospin) states.
The quantities $S_\alpha$ ($T$) and $s^i$
($t^i$) are the spin (isospin) of the pair  $j$, $k$ and of the third
particle $i$, respectively, and they are coupled to give
$\Sigma_\alpha$ ($\tauv$). 
The index  $\alpha$ stands for the set of  quantum numbers
\be\label{eq:acha}
  \a\equiv\{ \ell_\alpha, L_\alpha, \Lambda_\alpha, S_\alpha,
  \Sigma_\alpha\} \ .
\ee
The indices $\{\a,T,\tauv\}$ distinguish the various angular--spin--isospin
``channels''. A generic three--body w.f. can be written as
\be\label{eq:app1}
   \Psi=\sum_{i=1}^3 \psi(i)\ ,
    \quad \psi(i)\equiv \psi(\x_i,\y_i)\ ,
\ee
where each amplitude $\psi(i)$ corresponds to the
$i$--rearrangement configuration of the three particles and it can
be decomposed in channels as
\be\label{eq:tau1}
      \psi(i)=
     \sum_{\a,T,\tauv}  F_{\a,T,\tauv}(x_i,y_i)  \sai\ .
\ee
The antisymmetry of $\Psi$ can be simply enforced by requiring 
that the amplitude $\psi(i)$ change sign under the exchange of
particles $j$ and $k$. This can be easily done by including 
only those channels having odd values of
$\ell_\a+S_\a+T$ in the sum of Eq.~(\ref{eq:tau1})

In the following, the projection on states with
a definite value of the third component of the isospin of
the three particles will be useful. Let us define
\be\label{eq:xi}
  \Xi_i\equiv |t^i_z=-1/2,t^j_z=1/2, t^k_z=1/2\rangle
  \ ,
\ee
as the isospin state where particle $i$ is the neutron. Therefore,
each amplitude $\psi(i)$ can be decomposed as
\be\label{eq:tau2}
  \psi(i)=\sum_{\nu=1}^3 \psi_\nu(i) \Xi_\nu\ ,
\ee
where $\psi_\nu(i)$ is the component of the amplitude $\psi(i)$ and
particle $\nu$ is the neutron. To expand the components $\psi_\nu(i)$
in spherical waves, it is convenient to use
\be\label{eq:ang2}
    \sbi=
    \Bigl \{    \bigl[ Y_{\ell_\a}(\hat x_i)
        Y_{L_\a}(\hat y_i) \bigr  ]_{\Lambda_\a}
       \bigl [ (s^j s^k)_{S_\a} s^i \bigr ]_{\Sigma_\a}
       \Bigr \}_{J J_z} \ ,
\ee
which are a complete set of  angular--spin functions. The functions
$\sai$ and $\sbi$ are related to each other by
\bea
   \saiu &=& \sbi \biggl[ {1\over \sqrt{3}} \Xi_i +
                          {1\over \sqrt{3}} \Xi_j +
                          {1\over \sqrt{3}} \Xi_k\biggr] \ ,
                     \nonumber \\
\noalign{\medskip}
   \said &=& \sbi \biggl[
                          -{1\over \sqrt{2}} \Xi_j +
                          {1\over \sqrt{2}} \Xi_k\biggr] \ ,
                     \label{eq:tau4} \\
\noalign{\medskip}
   \sait &=& \sbi \biggl[ {2\over \sqrt{6}} \Xi_i -
                          {1\over \sqrt{6}} \Xi_j -
                          {1\over \sqrt{6}} \Xi_k\biggr] \ .
                     \nonumber
\eea
The components $\psi_\nu(i)$ can be then expanded as the sum
\be\label{tau3}
   \psi_\nu(i)=\sum_\b \widetilde F_{\b,\nu}(x_i,y_i)
   \sbi ,
\ee
where $\widetilde F_{\b,\nu}$ are radial amplitudes. Using
Eqs.~(\ref{eq:tau4}), it is not difficult to find the relation between
the amplitude $F_{\a,T,\tauv}$ defined in
Eq.~(\ref{eq:tau1}) and the functions $\widetilde F_{\b,\nu}$.

The $N-d$ system in the asymptotic region, where a nucleon  and the
deuteron are far apart, is described by the  following ``surface''
function,
\be\label{eq:suri}
  \Phi^d_{LS}(i)=
        \Bigl\{\bigl[
        \phi_d({\bf x}_i) s^i \bigr  ]_{S}  Y_{L}(\hat y_i)
      \Bigr \}_{J J_z} \bigl[ \xi_d t^i \bigr
       ]_{\tauv\tauv_z}\ ,\qquad \tauv={1\over 2}\ ,
\ee
where $\phi_d$ ($\xi_d$)  is the spatial--spin (isospin) part of
the deuteron w.f. The quantum number $L$ is the relative
angular momentum
between the lone nucleon and the deuteron, $S$ is the spin
obtained by coupling the spin $j= 1$ of the deuteron
with the spin 1/2 of the third nucleon.
The deuteron wave function
$\psi_d=\phi_d(\x_i)\times\xi_d$ satisfies the following
equation
\be\label{eq:deut}
  \left(-\htm \nabla^2_{x_i}
   + V^{\rm NN}(\x_i) \right) \psi_d= -B_d  \psi_d\ .
\ee

\section{The Asymptotic Behavior}
\label{sec:asy}

The asymptotic behavior of the w.f. describing $N-d$ scattering has
been discussed by many authors, in particular for the $n-d$
case~\cite{N72,G74,MGL76,GP92,PGF99}. 
The case of the scattering of three
particles, with at least two of them charged, has been considered in
Refs.~\cite{DN71,M80,AM93,FP96}. In this section, we rederive the 
asymptotic behavior of the w.f. for a $p-d$ scattering process, in
order to clarify some properties used to
discuss the KVP in the following section. 


The asymptotic behavior of the amplitudes $\psi(i)$ given in
Eq.~(\ref{eq:app1}) is analyzed by considering them as 
solutions of the three Faddeev--Noble equations (FNE)~\cite{N67}
\be\label{eq:bpp2}
   \left[ -\htm \biggl( \nabla^2_{x_i} + \nabla^2_{y_i} \biggr)
   + V_C + V^{\rm NN}(\x_i) -E \right] \psi(i) =-V^{\rm NN}(\x_i)
   \biggl(\psi(j)+\psi(k)\biggr)
   \ ,
\ee
with $i,j,k$ cyclic.
The 3N interaction $W^{3N}$ has been disregarded since we are
interested in the asymptotic solutions of Eq.~(\ref{eq:bpp2}), where
at least one particle is very far from the other two.
Let us consider the volume ${\cal V}(R)$ of the six--dimensional space
where $\rho\ge R$ ($R\rightarrow\infty$). ${\cal V}(R)$ can be divided
into the following
regions,  corresponding to different ranges of values of the
hyperangular variables $\Omega$.

\begin{itemize}

\item{\it 1+1+1 (breakup) region.} Here all the
 particles are well separated from each other. This breakup region
 is denoted as
 ${\cal V}_b(R)\equiv\{\rho\ge R; x_i\ra\infty\ {\rm for}\
 i=1,2,3\}$.

\item{\it 2+1 regions.}
The part of ${\cal V}(R)$ where the particles $j$ and $k$ are close
enough to interact through the NN potential, while particle $i$
is very far from them, is hereafter denoted by ${\cal V}_i(R)$.
There are three such regions, corresponding to $i=1,2,3$, namely
${\cal V}_i(R)\equiv\{\rho\ge R; x_i< a_N\}$. In ${\cal V}_i(R)$,
we have $y_i \stackrel{\sim}{>} R$ but the distance $x_i$ between
particles $j$ and $k$ is of the order or lesser than the range
$a_N$ of the nuclear potential. As $R$ increases, the range of the
hyperangle $\hypfi_i$ subtended by the region ${\cal V}_i(R)$
decreases, and in the limit $R\ra\infty$ it reduces virtually
only to the value $\hypfi_i=\pi/2$.

\item{\it Transitions regions.}
The regions between ${\cal V}_b(R)$ and ${\cal V}_i(R)$ are
denoted as ${\cal V}_{t,i}(R)$, $i=1,2,3$. They can be defined as
${\cal V}_{t,i}(R)\equiv \{\rho\ge R; x_i\le
\Gamma \rho^\mu;x_i>a_N\}$, where $\Gamma$ and $\mu$ are suitable
constants ($0<\mu<1$, as will be shown below). In this region
\be\label{eq:pd8}
  \cos\hypfi_i = {x_i\over \rho} \le
   \Gamma \rho^{\mu-1}\ra 0\ , {\rm for \ } \rho
   \ra\infty\ .
\ee
Therefore, for $\rho\ra\infty$ the hyperangle extension of this
region reduces again to the value $\hypfi_i=\pi/2$. Such regions
were considered for the first time by Merkuriev~\cite{M74,M80}
and should allow for a smooth transition between the solutions in
the regions ${\cal V}_b(R)$ and ${\cal V}_i(R)$.

\end{itemize}

Let us study the asymptotic behavior of the solution of the FNE
given in Eq.~(\ref{eq:bpp2}) in the above described regions.

\subsection{The ${\cal V}_b(R)$ region}
\label{sec:Vb}

In the breakup region ${\cal V}_b(R)$ the nuclear interaction can
be safely disregarded and the FNE for $\psi(i)$ becomes
\be\label{eq:bpp3}
   \left[ -\htm \biggl( \nabla^2_{x_i} + \nabla^2_{y_i} \biggr)
   + V_C  -E \right] \psi(i) = 0
   \ .
\ee
By rewriting this equation using the $\rho,\Omega_i$ coordinates
and disregarding terms  vanishing faster than
${\cal O}(1/\rho)$ as $\rho\ra\infty$, the solution reduces to~\cite{M80}
\bea
  \psi(i)&=&
     {  e^{\ii Q\rho -\ii  \hat\chi\ln 2 Q\rho} \over \rho^{5\over 2}}
   \left (  \sum_{\a,T,\tauv}  A_{\a,T,\tauv}(\hypfi_i)\sai \right )
    \nonumber \\
\noalign{\medskip}
   && \qquad
   + {  e^{-\ii Q\rho +\ii  \hat \chi\ln 2 Q\rho} \over \rho^{5\over 2}}
   \left (  \sum_{\a,T,\tauv}   B_{\a,T,\tauv}(\hypfi_i)\sai\right )
   \ ,\quad {\rm region}\ {\cal V}_b(R)\ ,\label{eq:pd4b}
\eea
where $ E=\hbar^2 Q^2/m $ and
\be
   \hat \chi = {m\over 2\hbar^2 Q}
  \sum_{i=1}^3 {e^2\over \cos\hypfi_i} {1+\tau_{j,z} \over 2}
  {1+\tau_{k,z} \over 2} \ , 
\label{eq:chi}
\ee
is a dimensionless operator. The functions
$A_{\a,T,\tauv}(\hypfi_i)$ and $B_{\a,T,\tauv}(\hypfi_i)$ are the
so--called breakup amplitudes and are 
fixed by the dynamics of the process.
For example, the cross section for the process $p+d\ra
p+p+n$ is proportional to $|A_{\a,T,\tauv}(\hypfi_i)|^2$.

Alternatively, $\psi(i)$ can be
decomposed as in Eq.~(\ref{eq:tau2}), where the neutron is explicitly
labelled. The corresponding components
$\psi_\nu(i)$ are
\bea
  \psi_\nu(i)&=&
     {  e^{\ii Q\rho -\ii
     \eta(\hypfi_\nu)\ln 2 Q\rho} \over \rho^{5\over 2}}
   \left (  \sum_\a  \widetilde A^i_{\a,\nu}(\hypfi_i)\sbi \right )
    \nonumber \\
\noalign{\medskip}
   && \qquad
   + {  e^{-\ii Q\rho +\ii \eta(\hypfi_\nu) \ln 2 Q\rho}
     \over \rho^{5\over 2}}
   \left (  \sum_\a   \widetilde B^i_{\a,\nu}(\hypfi_i)\sbi\right )
   \ ,\quad {\rm region}\ {\cal V}_b(R)\ ,\label{eq:pd4c}
\eea
where $\eta(\hypfi_\nu)$ represents the isospin matrix elements of the
operator $\hat \chi$ and is defined as
\be\label{eq:etai}
  \eta(\hypfi_\nu)={me^2\over 2\hbar^2Q\cos\hypfi_\nu} \ ,
\quad \nu=1,2,3\ .
\ee
The breakup amplitudes depending on the neutron label $\nu$ are related
to the previous ones by the relations:
\bea
   \widetilde A^i_{\b,i}(\hypfi_i) &=&
                       {1\over \sqrt{3}} A_{\b,1,{3\over 2}}(\hypfi_i)
                     + {2\over \sqrt{6}} A_{\b,1,{1\over 2}}(\hypfi_i) \ ,
                     \nonumber \\
\noalign{\medskip}
   \widetilde A^i_{\b,j}(\hypfi_i) &=&
                       {1\over \sqrt{3}} A_{\b,1,{3\over 2}}(\hypfi_i)
                     - {1\over \sqrt{6}} A_{\b,1,{1\over 2}}(\hypfi_i)
                     - {1\over \sqrt{2}} A_{\b,0,{1\over 2}}(\hypfi_i)
                      \ ,
                     \label{eq:tau5} \\
\noalign{\medskip}
   \widetilde A^i_{\b,k}(\hypfi_i) &=&
                       {1\over \sqrt{3}} A_{\b,1,{3\over 2}}(\hypfi_i)
                     - {1\over \sqrt{6}} A_{\b,1,{1\over 2}}(\hypfi_i)
                     + {1\over \sqrt{2}} A_{\b,0,{1\over 2}}(\hypfi_i) \ ,
                     \nonumber
\eea
and analogously for $\widetilde B^i_{\a,i}$. The superscript $i$ of
the functions $\widetilde A^i_{\b,\nu}$ recalls that these amplitudes are the
appropriate combinations of functions $A_{\a,T,\tauv}$ entering the
expression~(\ref{eq:pd4c}) of $\psi_\nu(i)$.

\subsection{The ${\cal V}_{t,i}(R)$ region}
\label{sec:Vip}

In the following analysis we will fix our attention to the term describing
outgoing breakup waves, i.e. proportional to $\exp(\ii Q\rho)$.
When  $\cos\hypfi_\nu\ra 0$, namely  going into the regions ${\cal
V}_{t,\nu}(R)$, the term $\exp(- \ii
\eta(\hypfi_\nu) \ln2Q\rho)$ in Eq.~(\ref{eq:pd4c}) oscillates increasingly. 
Moreover, it is no longer appropriate to
disregard terms like $1/ (\rho \cos\hypfi_\nu)^2$ in the construction
of the solution.
In this region the FNE satisfied 
by $\psi(i)$ is still given by Eq.~(\ref{eq:bpp3}), since $x_i> a_N$. 
We can distinguish the following cases.

\begin{figure}
\psfig{file=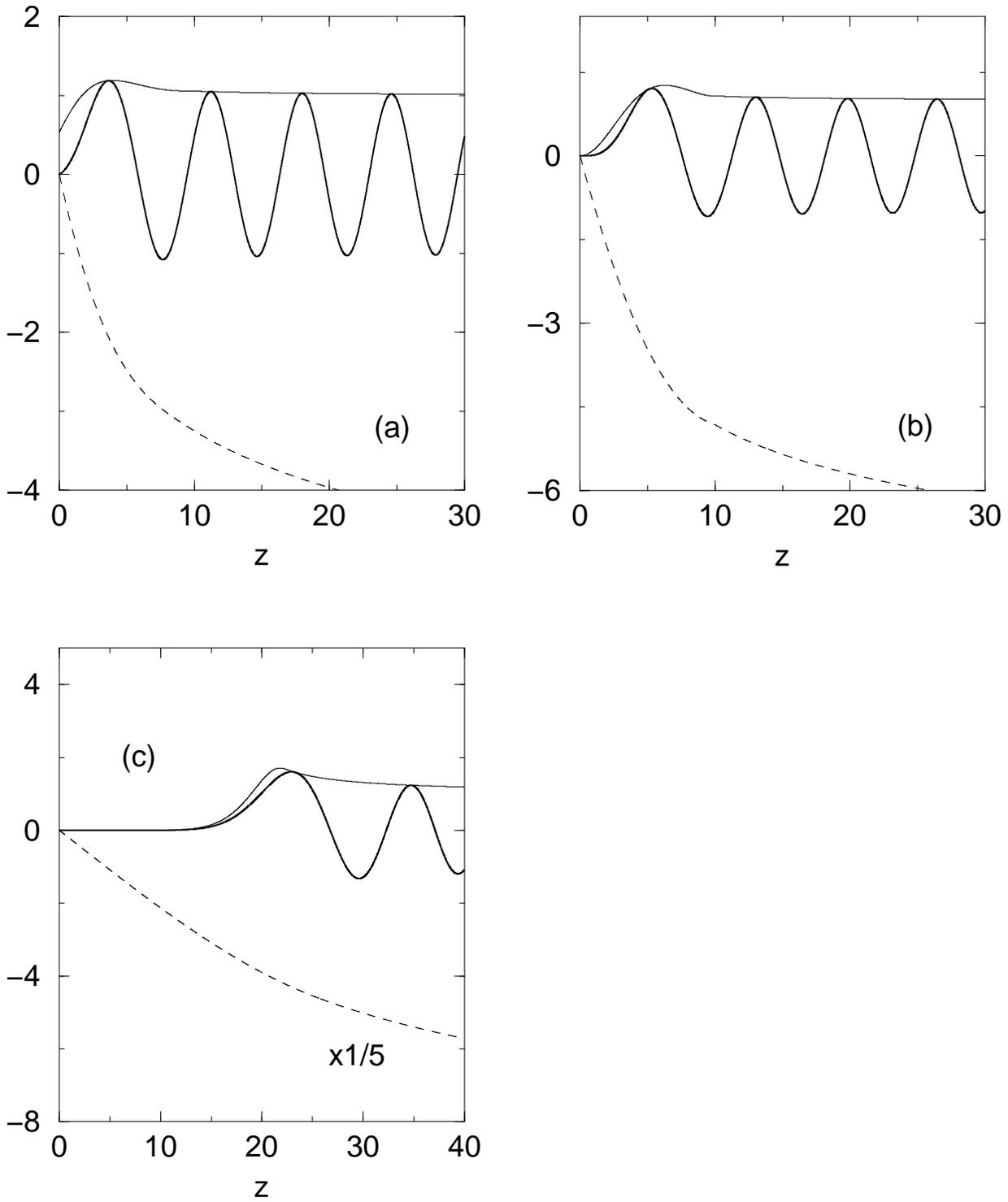,height=20cm}
\label{fg:fdec}
\caption{Decomposition of the regular Coulomb function $F_\ell(\eta,z)
= {\cal F}_\ell(\eta,z)\sin[z+ \beta_\ell(\eta,z)]$ for a few values
of $\eta$ and $\ell$. Case (a): $\ell=0$ and $\eta=1$;
Case (b): $\ell=2$ and $\eta=1$; Case (c): $\ell=0$ and $\eta=10$.
The thick solid, thin solid and the dashed lines show the functions
$F_\ell(\eta,z)$, ${\cal F}_\ell(\eta,z)$ and $\beta_\ell(\eta,z)$,
respectively. }
\end{figure}

\noindent 1. Component $\psi_{i}(i)$ (particle $i$ is a neutron). 
The FNE satisfied by this component
  is
\be
   \left[ -\htm \biggl( \nabla^2_{x_i} + \nabla^2_{y_i} \biggr)
   + {e^2\over x_i}  -E \right] \psi_i(i) = 0
   \ .\label{eq:bpp4}
\ee
  Since we are in a region where
  $x_i\ll y_i\approx \rho$,
  the general solution of the above equation can be written as
 \be
  \psi_{i}(i)= \sum_\b \int_0^{\pi\over 2} d\varphi \;
  {\cal C}_{\b,i}(\varphi)
  {1\over x_i} F_{\ell_\b}\biggl( {me^2\over
    2\hbar^2 q},q x_i\biggr)
  {1\over y_i}\; \hat h^+_{L_\b}(p y_i)\; \sbi \ .
  \label{eq:pp1}
\ee
where
\be\label{eq:pp2}
 q=Q\cos\varphi\ ,\qquad  p=Q\sin\varphi\ ,
\ee
$F_{\ell}(\eta,qx)$  is the regular Coulomb function and
\be\label{eq:hpiu}
  \hat h^+_{L}(z)= z \Bigl ( -y_{L}(z) + \ii j_{L}(z) \Bigr )
  \ ,\quad \hat h^+_{L}(z)\ra e^{\ii (z-L\pi/2)} \ {\rm for}
  \ z\ra\infty\ ,
\ee
with $j_L$ and $y_L$ spherical Bessel functions.
The function ${\cal C}_{\b,i}(\varphi)$ entering Eq.~({\ref{eq:pp1}),
whose dependence on $Q$  is implicit, is a 
smooth weight function, strictly related to the breakup amplitude $
A_{\b,T,\tauv}(\hypfi_i)$. It goes to zero for $\varphi\ra0$ or
$\pi/2$. In order to analyze the asymptotic form of the solution
it is convenient to decompose the Coulomb function as
\be\label{eq:regf}
  F_\ell(\eta,z)= {\cal F}_\ell(\eta,z)
   \sin\Bigl(z+\beta_\ell(\eta,z)\Bigr)\ ,
\ee
where ${\cal F}_\ell(\eta,z)$ and $\beta_\ell(\eta,z)$ have the
following asymptotic behaviors for $z\gg \max(1,\eta^2)$,
\bea
  {\cal F}_\ell(\eta,z)&=& 1\ , \label{eq:regf1}\\
  \beta_\ell(\eta,z)&=& -\eta\ln 2z-\ell\pi/2
  +\sigma_\ell(q)\ ,
  \label{eq:regf2}
\eea
$\sigma_\ell(q)$ being the Coulomb phase--shift. For
$z\ll\min(1,1/\eta)$ we have instead
\bea
  {\cal F}_\ell(\eta,z)&=& {2^\ell e^{-\pi\eta/2}
    |\Gamma(\ell+1+\ii\eta)| \over \Gamma(2\ell+2) } (z)^\ell
     \label{eq:regf3}\\
  \beta_\ell(\eta,z)&=& 0\ ,
  \label{eq:regf4}
\eea
where $\Gamma$ is the gamma function. The
decomposition~(\ref{eq:regf}) is somewhat arbitrary. However, we
require that both ${\cal F}$ and $\beta$ always be finite.
Thus, $z_n+\beta_\ell(\eta,z_n)=n\pi$ for $n=0,1,\ldots$, where $z_n$
are the zeros of the function $F_\ell(\eta,z)$. An example of
the decomposition is shown in Fig.~(\ref{fg:fdec}) for a few values of
$\eta$ and $\ell$.

In the large $\rho$ limit the integral in Eq.~(\ref{eq:pp1}) can
be evaluated by the saddle--point approximation. Introducing the above
decomposition for the Coulomb function and the relations
$x_i=\rho\cos\hypfi_i$ and $y_i=\rho\sin\hypfi_i$,
the leading term, given by the saddle--point $\varphi=\hypfi_i$, is
\bea
  \psi_{i}(i)&=& \sum_\b \sqrt{2\pi\over Q} e^{3\pi \ii /4}
    {{ \cal C}_{\b,i}(\hypfi_i) \over
    2\ii\;\cos\hypfi_i\sin\hypfi_i}
   { \cal F}_{\ell_\b}\biggl( \eta(\hypfi_i)
       ,Q\rho \cos^2\hypfi_i\biggr)
     \nonumber \\
   && \qquad\qquad \times
   \exp\biggl[\ii \beta_{\ell_\b}\Bigl(  \eta(\hypfi_i)
    ,Q\rho \cos^2\hypfi_i\Bigr)
    -\ii L_\b{\pi\over 2}\biggr]
   \; {e^{\ii Q\rho} \over \rho^{5\over2}}
     \sbi \nonumber \\
\noalign{\medskip}
   &=&
     \sum_\b    { \cal F}_{\ell_\b}\biggl(  \eta(\hypfi_i)
    ,Q\rho \cos^2\hypfi_i\biggr) \; \nonumber \\
   &&
      \times \exp\biggl[\ii \beta_{\ell_\b}\Bigl(  \eta(\hypfi_i)
    ,Q\rho \cos^2\hypfi_i\Bigr)
    +\ii  \eta(\hypfi_i) \ln(2Q\rho \cos^2\hypfi_i)
    +\ii \ell_\b{\pi\over 2} -\ii \sigma_{\ell_\b}(Q\cos\hypfi_i)\biggr]
      \nonumber \\
   && \times \widetilde
     A^i_{\b,i}(\hypfi_i) \; {1\over\rho^{5\over2}} \;
    \exp\biggl[\ii Q\rho- \ii \eta(\hypfi_i)
      \ln(2Q\rho)\biggr] \sbi\ ,      \ .\label{eq:exc1}
\eea
where in the last passage we have defined
\bea\label{eq:exc3}
 \widetilde A^i_{\b,i}(\hypfi_i) &=&  \sqrt{2\pi\over Q} e^{3\pi \ii /4}
    {{ \cal C}_{\b,i}(\hypfi_i) \over
    2\ii\;\cos\hypfi_i\sin\hypfi_i}  \times \nonumber \\
   &&   \exp\biggl( - \ii  \eta(\hypfi_i)
        \ln  (\cos^2\hypfi_i)
    -\ii (\ell_\b+L_\b) {\pi\over 2}+\ii \sigma_{\ell_\b}(Q\cos\hypfi_i)
    \biggr )\ .
\eea
The next term of the saddle--point expansion is $\sim
\rho^{-7/2}$. In the region ${\cal V}_{t,i}(R)$ we have $\cos\hypfi_i\ra0$
so that $\eta^2(\hypfi_i)\gg1$. Therefore, if
\be\label{eq:exc2}
    Q\rho \cos^2\hypfi_i \gg  \eta^2(\hypfi_i)
    \ ,
\ee
or equivalently
\begin{equation}\label{eq:vpi}
  x_i \gg \rho_c\equiv \left (m^2e^4\rho^3\over 4\hbar^4Q^3
   \right )^{1\over 4}\ ,
\end{equation}
the functions ${\cal F}$ and $\beta$ reach their asymptotic values
given in Eqs.~(\ref{eq:regf1}) and~(\ref{eq:regf2}). Then,
$\psi_{i}(i)$ reduces to the form of Eq.~(\ref{eq:pd4c}) with
$\widetilde A^i_{\b,i}$ given by Eq.~(\ref{eq:exc3}). When
Eq.~(\ref{eq:vpi}) is not satisfied, corresponding to the situation in
which the two protons approach each other, the functions ${\cal F}$ and
$\beta$ start deviating from their asymptotic  behavior 
and the breakup w.f.  cannot be cast into the form~(\ref{eq:pd4c}) 
and~(\ref{eq:pd4b}). The extreme situation corresponds to the condition
\be\label{eq:exc4}
    Q\rho \cos^2\hypfi_i \ll   { 1\over  \eta(\hypfi_i)}
    \ ,
\ee
or equivalently,
\be\label{eq:exc5}
    x_i\ll \rho_c'\equiv {2\hbar^2\over me^2}\approx 60\;{\rm fm}
    \ ,
\ee
and the functions ${\cal F}$ and $\beta$ reach the behavior given in
Eqs.~(\ref{eq:regf3}) and (\ref{eq:regf4}). Thus,  ${\cal F}\ra
e^{-\pi me^2/\hbar^2Q\cos\hypfi_i}\ra 0$ since
$x_i/\rho=\cos\hypfi_i\ra0$. Also the successive terms in the
saddle--point expansion are zero. This corresponds to
the intuitive fact that the probability of finding two protons (in
this case, labelled as particles $j$ and $k$) at any finite
distance $x_i= r_{jk}$ is negligible in the asymptotic region
$R\ra\infty$. We arrive at the important result that 
the amplitude $\psi_{i}(i)$  goes smoothly
from the behavior given in Eq.~(\ref{eq:pd4c}) to a vanishing
value when going from the region where $x_i\gg \rho_c$ to the one
where $x_i\ll \rho_{c}'$.

The same conclusion holds if the regular Coulomb function
$F_{\ell}(\eta,z)$ is replaced by the (regular) solution
of the two--body Schroedinger equation
\be\label{eq:fv}
 \left [ {d^2\over dz^2} + 1 -{2\eta\over z}-{\ell(\ell+1)\over z^2}
  -{mV^{\rm NN}(x_i)\over \hbar^2q^2}\right ]
  F^{\rm V}_\ell(\eta,q,x_i) = 0\ ,\quad z=qx_i\ ,
\ee
where,  for the matter of simplicity,
we have considered here a central NN potential $V^{\rm NN}(x_i)$.
Outside the range of $V^{\rm NN}$ we have
$F^{\rm V}_\ell\ra F_\ell + \tan\delta_\ell \; G_\ell$ where $G$ is the
irregular Coulomb function and $\delta_\ell$ the
phase shift induced by the NN potential. However, for $q\ra
0$ or $\eta\ra\infty$ it 
can be shown that $G\ra \exp(\pi\eta)$ and $\tan\delta_\ell\ra
\exp(-2\pi\eta)$. Therefore, when $x_i\ll \rho_c'$ again $F^{\rm V}\ra
0$. }

\noindent 2. Component $\psi_{j}(i)$ (the neutron is labelled as $j$).
  The equation satisfied by  $\psi_j(i)$ in this region is
\be\label{eq:pn0}
   \left[ -\htm \biggl( \nabla^2_{x_i} + \nabla^2_{y_i} \biggr)
   + {e^2\over x_j} -E \right] \psi_j(i) = 0
   \ .
\ee
  In ${\cal V}_{t,i}(R)$, $x_j\approx \kappa y_i$ and therefore
the solution of the above equation can be written in the form
 \bea
  \psi_{j}(i)&=& \sum_\b \int_0^{\pi\over 2} d\varphi \;
  {\cal C}_{\b,j}(\varphi)
   {1\over x_i} \hat j_{\ell_\b}(q x_i) \nonumber \\
  && \qquad
   {1\over y_i} \;\Bigl [  G_{L_\b}({me^2\over 2\hbar^2 \kappa p},p y_i)+ \ii
    F_{L_\b}({me^2\over 2\hbar^2 \kappa p},p y_i)\Bigr]  \; \sbi \ .
  \label{eq:pn1}
 \eea
where $G_L$ is the irregular Coulomb function and $\hat
j_\ell$ is a regular Riccati-Bessel function. As has been done 
before, this latter function can be decomposed as,

\be\label{eq:pn2}
  \hat  j_\ell(z)={\cal J}_\ell(z) \sin(z+\zeta_\ell(z))\ , 
\ee
where ${\cal J}_\ell(z)$ and $\zeta_\ell(z)$ have the following
asymptotic behavior for $z\gg 1$,
\bea
  {\cal J}_\ell(z)&=& 1\ , \label{eq:regf1n}\\
  \zeta_\ell(z)&=& -\ell\pi/2\ ,  \label{eq:regf2n}
\eea
and for $z\ll 1$,
\bea
  {\cal J}_\ell(z)&=& {(z)^\ell\over (2\ell+1)!!}
                                \ , \label{eq:regf3n}\\
  \zeta_\ell(z)&=& 0\ .  \label{eq:regf4n}
\eea
Note that ${\cal J}_\ell(z)= {\cal F}_\ell(0,z)$ and
$\zeta_\ell(z)= \beta_\ell(0,z)$. Moreover,
${\cal J}_0(z)= 1$ and $\zeta_0(z)= 0$.

Following a procedure similar to the one adopted in point 1), 
the following expression for the component
$\psi_j(i)$ in ${\cal V}_{t,i}(R)$  is found,
\be\label{eq:pn3}
  \psi_{j}(i)= C(\rho)
     \sum_\b {\cal J}_{\ell_\b}(Q\cos\hypfi_i x_i)\;
     \exp\biggl[\ii \zeta_{\ell_\b}(Q\cos\hypfi_i x_i) + \ii \ell_\b
        {\pi\over 2}
         \biggr]  \widetilde A^i_{\b,j}(\hypfi_i) \sbi \ , 
\ee
where
\be\label{eq:pn4}
 \widetilde A^i_{\b,j}(\hypfi_i) =  \sqrt{2\pi\over Q} e^{3\pi \ii /4}
    {{ \cal C}_{\b,j}(\hypfi_i) \over
    2\ii\;\cos\hypfi_i\sin\hypfi_i}
     \exp\biggl( -\ii (\ell_\b+L_\b) {\pi\over 2}+
        \ii \sigma_{\ell_\b}(Q\sin\hypfi_i)
    \biggr )\ .
\ee
In Eq.~(\ref{eq:pn3}), we have defined
\be\label{eq:ee}
   \quad C(\rho)= {1\over \rho^{5\over 2}}
    \exp\left(\ii Q\rho -\ii  {me^2\over 2\hbar^2 Q \k}
        \ln 2 Q\rho\right) \ . 
\ee
Therefore, for
\begin{equation}\label{eq:vpi2}
  x_i \gg  \rho_n\equiv \left ({\rho\over Q} \right )^{1\over 2}\ ,
\end{equation}
where the behavior given in Eqs.~(\ref{eq:regf1n})
and~(\ref{eq:regf2n}) for ${\cal J}_\ell(qx_i)$ and
$\zeta_\ell(qx_i)$ is valid, the component $\psi_j(i)$
coincides with 
Eq.~(\ref{eq:pd4c}) [note that $\cos\hypfi_j\approx \kappa$  in ${\cal
V}_{t,i}(R)$].

If $x_i\ll \rho_n$, the amplitudes of the channels having $\ell_\b>0$
become vanishing small since ${\cal J}_{\ell_\b}\propto
(x_i/\rho_n)^{2\ell_\b}$. This corresponds to the intuitive fact
that the probability of finding two nucleons (in this case,
labelled as particles $j$ and $k$) with a vanishing relative
velocity and relative orbital angular momentum greater than zero is
negligible in the asymptotic region $R\ra\infty$. Therefore, 
\be\label{eq:pn5}
    \psi_j(i)\ra C(\rho) \sum_\a \delta_{\ell_a,0}
     \widetilde A^i_{\a,j}({\pi\over 2})
    \sbi \ ,\qquad x_i\ll\rho_n \ .
\ee

\noindent 3. Component $\psi_{k}(i)$ (the neutron is labelled as $k$).
  The solution for this case coincides with that  discussed in point
  2.

Finally, we can estimate of the value of the constant $\mu$
defining the region ${\cal V}_{t,i}(R)$. In the $n-d$ case, the
appropriate value 
is $\mu\approx 1/2$ [see Eq.~(\ref{eq:vpi2})].  In the $p-d$ case, a
value of  $\mu \approx 3/4$ can be chosen in accordance with
Eq.~(\ref{eq:vpi}). As shown in
Ref.~\cite{M74}, the presence of those regions does not alter the
validity of the KVP for the $n-d$ case. Therefore, it can be expected
that such regions do not play any important role also in the $p-d$
case, as will become evident in the next section.

\subsection{The ${\cal V}_i(R)$ region}
\label{sec:Vi}

In this region the nuclear interaction between particles $j$ and $k$ cannot be
disregarded, whereas particle $i$ is far from the other two. The
solution consists of an elastic
$p-d$ amplitude $\psi^{(e)}(i)$ plus a breakup amplitude
$\psi^{(b)}(i)$, namely $\psi(i)=\psi^{(e)}(i)+\psi^{(b)}(i)$. The amplitude
$\psi^{(e)}$ has the form
\be
  \psi^{(e)}(i) = \sum_{LS} \Phi^d_{LS}(i)
       \biggl[ a_{LS} {\cal H}^+_L(\eta_0,q_0\d_i)
             + b_{LS} {\cal H}^-_L(\eta_0,q_0\d_i)
    \biggr] \ , \label{eq:pd9}
\ee
where $\Phi^d_{LS}(i)$ is given in Eq.~(\ref{eq:suri}), 
the distance $d_i$ in Eq.~(\ref{eq:dk}) and
\be\label{eq:coul}
  {\cal H}_L^\pm(\eta_0,q_0 \d_i)=
{ G_L(\eta_0,q_0 \d_i)\pm \ii F_L(\eta_0,q_0 \d_i) \over q_0 \d_i}\ .
\ee
The functions $F$ and $G$ are the
regular and irregular Coulomb functions and
\be\label{eq:pd10}
  \eta_0= {me^2\over 2\hbar^2 q_0}\ ,
\ee
is the usual Coulomb parameter, with the wave number $q_0$ satisfying
the relation
\be\label{eq:preg1}
  -B_d+ {1\over (\k)^2} \htm q_0^2 = E\ .
\ee
The coefficients $a_{LS}$ ($b_{LS}$)
are the amplitudes of the corresponding  incoming
(outgoing) waves. In fact, ${\cal H}^\pm \approx \exp(\pm \ii
 q_0 d_i + \ldots)$
in the limit of large $q_0d_i$ values.

The amplitude $\psi^{(b)}(i)$, when particles $j$ and $k$ are protons,
again vanishes for $\rho\ra\infty$ due to the Coulomb
repulsion, namely $\psi^{(b)}_i(i)=0$.
When one of the particles is a neutron, it takes the
form
\be\label{eq:pd11}
   \psi^{(b)}_{\nu}(i)= \sum_\b \delta_{\ell_\b,0}\;
   \phi^0_{\b}(x_i) \;\widetilde A^i_{\b,\nu}({\pi\over2})
    \; C(\rho)\;
   \sbi \ ,\qquad \nu=j\ {\rm or}\ k\ ,
\ee
where $\phi^0_{\a}$ is the s--wave solution of a
inhomogeneous zero energy two--body Schroedinger--like
equation~\cite{FP96,PGF99} derived from 
Eq.~(\ref{eq:bpp2}) by taking into account the coupling between
$\psi(i)$ and $\psi(j)+\psi(k)$
induced by the NN potential. The amplitude $\psi^{(b)}_{j\ {\rm or}\
k}(i)$ vanishes for  $\ell_\b>0$ due to the centrifugal repulsion.
For the present discussion of the KVP the
precise form of the functions $\phi^0_{\b}$ is not
important. They are normalized so that $\phi^0_{\a}\ra 1$
outside ${\cal V}_i(R)$, namely when $x_i\gg a_N$, in order to match
the result obtained in the transition region.
The derivation of the equation satisfied by $\phi^0_\a$ is reported in
the Appendix. 

\subsection{The ${\cal V}_j(R)$+${\cal V}_{t,j}(R)$ regions}
\label{sec:Vj}

In these regions $y_j\ra\infty$ and $x_j/y_j\ra 0$ and therefore
we can approximate $x_i\sim\kappa y_j$ and $y_i\sim
y_j/2$.  Then, the component $\psi_i(i)$
satisfies again the uncoupled FNE given in
Eq.~(\ref{eq:bpp3}).
Moreover, $\cos \hypfi_i\sim \kappa$ or $\hypfi_i\sim\pi/6$ and from
Eq.~(\ref{eq:exc1})
\be\label{eq:vj1}
  \psi_i(i)\ra C(\rho) \sum_\a \widetilde A^i_{\a,i}({\pi\over 6})
  \sbi\ ,\qquad {\rm region}\ {\cal V}_j(R)\ .
\ee
In fact, here (for $\rho\ra\infty$) we always have $Q\rho\cos^2\hypfi_i\gg
\eta^2(\hypfi_i)$. The same conclusion is reached for
$\psi_k(i)$. Both components can be obtained directly by
taking the limit $\hypfi_i\ra\pi/6 $ in Eq.~(\ref{eq:pd4c}).

The component $\psi_j(i)$ is more difficult to obtain. In fact,
since here $x_j$ can become very small,
we have to fully take into account the distortion due to the Coulomb
potential. 
This component can be calculated by expressing it in terms of
the coordinates $\x_j$ and $\y_j$ and by rewriting
Eq.~(\ref{eq:bpp3}) as
\be
   \left[ -\htm \biggl( \nabla^2_{x_j} + \nabla^2_{y_j} \biggr)
   + {e^2\over x_j}  -E \right] \psi_j(i;\x_j,\y_j) = 0
   \ .\label{eq:bpp5}
\ee
Now $\psi_j(i;\x_j,\y_j)$ can be determined by following the same
procedure as in Sect.~\ref{sec:Vip}. In particular, it has been shown
there that for $x_j\ll \rho_c'$, $\psi_j(i)$ vanishes due to the Coulomb
repulsion between the protons.

The behavior of $\psi(i)$ in ${\cal V}_k(R)$ and ${\cal V}_{t,k}(R)$ can
be analyzed in the same manner. Here the components $\psi_i(i)$
and $\psi_j(i)$ are given by Eq.~(\ref{eq:vj1}) and can be obtained
directly as the limit of the solution valid in ${\cal V}_b(R)$. The
component $\psi_k(i)$ will be distorted by the Coulomb repulsion and
will vanish inside ${\cal V}_k(R)$.

\subsection{Summary}
\label{sec:summ}

It is straightforward to show that the behavior of the
incoming wave $\propto \exp(-\ii Q\rho+\ldots)$ has
similar properties. Therefore, the behavior of the amplitudes
$\psi(i)$ in the asymptotic regions can be put in the form,
\bea
  \psi(i) &=& \sum_{LS} \Phi^d_{LS}(i)
       \biggl[ a_{LS}
        {\cal H}^+_L(\eta_0,q_0\d_i)
       + b_{LS}  {\cal H}^-_L(\eta_0,q_0\d_i)
        \biggr] \nonumber \\
  && +   {  e^{\ii Q\rho -\ii \hat \chi \ln(2Q\rho) } \over \rho^{5\over 2}}
   \left (  \sum_{\a,T,\tauv}
    {\cal A}_{\a,T,\tauv}(\rho,\Omega_i)\sai\right )
    \nonumber \\
\noalign{\medskip}
   && \qquad
   + {  e^{-\ii Q\rho +\ii \hat \chi\ln 2 Q\rho} \over \rho^{5\over 2}}
   \left (  \sum_{\a,T,\tauv}
    {\cal B}_{\a,T,\tauv}(\rho,\Omega_i)\sai\right )
   \ .\label{eq:pd4}
\eea
The ``distorted'' breakup amplitudes ${\cal A}_{\a,T,\tauv}$
and ${\cal B}_{\a,T,\tauv}$
have been written as functions of ($\rho$, $\Omega_i$) taking into
account the modifications induced by the Coulomb, centrifugal and NN
potentials.
In ${\cal V}_b(R)$, clearly ${\cal A}_{\a,T,\tauv}(\rho,\Omega_i)\ra
A_{\a,T,\tauv}(\hypfi_i)$
and ${\cal B}_{\a,T,\tauv}(\rho,\Omega_i)\ra B_{\a,T,\tauv}(\hypfi_i)$.
Moreover, as discussed in Sect.~\ref{sec:Vip} and~\ref{sec:Vi}, in the
region ${\cal V}_i(R)$ and ${\cal V}_{t,i}(R)$, a good approximation for
the (isospin-projected components of) ${\cal A}_{\a,T,\tauv}(\rho,\Omega_i)$
should be
\bea
 \widetilde {\cal A}^i_{\b,i}(\rho,\Omega_i)&\ra&
      {\cal F}^{(c)}_{\ell_\b}(x_i,\hypfi_i)
     \widetilde A^i_{\b,i}(\hypfi_i)\ ,\label{eq:vb1}     \\
 \widetilde {\cal A}^i_{\b,j}(\rho,\Omega_i)&\ra&
      {\cal F}^{(n)}_{\ell_\b}(x_i,\hypfi_i)
     \widetilde A^i_{\b,j}(\hypfi_i)\ ,\label{eq:vb2} \\
 \widetilde {\cal A}^i_{\b,k}(\rho,\Omega_i)&\ra&
      {\cal F}^{(n)}_{\ell_\b}(x_i,\hypfi_i)
     \widetilde A^i_{\b,k}(\hypfi_i)\ ,\label{eq:vb3}
\eea
where the following functions have been introduced,
\bea
     {\cal F}^{(c)}_{\ell_\b}(x_i,\hypfi_i) &=&
     {\cal F}_{\ell_\b}\Bigl(  \eta(\hypfi_i)
                  ,Q\rho \cos^2\hypfi_i\Bigr)
     \exp\biggl[\ii \beta_{\ell_\b}\Bigl(  \eta(\hypfi_i)
    ,Q\rho \cos^2\hypfi_i\Bigr) \nonumber \\
    && \qquad\qquad\qquad
      +\ii  \eta(\hypfi_i) \ln(2Q\rho\cos^2\hypfi_i)
      +\ii \ell_\b{\pi\over 2} -\ii \sigma_{\ell_\b}(Q\cos\hypfi_i)\biggr]
      \ ,  \label{eq:vb4} \\
\noalign{\medskip}
   {\cal F}^{(n)}_{\ell_\b}(x_i,\hypfi_i) &=&
      \Bigl(\delta_{{\ell_\b},0} (\phi^0_{\b}(x_i)-1) + 1\Bigl)\;
      {\cal J}_{\ell_\b}\Bigl( Q\rho \cos^2\hypfi_i\Bigr)
       \nonumber \\
      && \qquad\qquad\qquad \times  \exp\biggl[\ii \zeta_{\ell_\b}
     \Bigl( Q\rho \cos^2\hypfi_i\Bigr)
    +\ii \ell_\b{\pi\over 2} \biggr]
      \ .  \label{eq:vb5}
\eea
In the regions ${\cal V}_i(R)+{\cal V}_{t,i}(R)$
the divergent behavior of $\hat \chi$ for $\cos\hypfi_i\ra 0$ is
completely canceled by terms in ${\cal A}_{\a,T,\tauv}$. 
It can be checked that outside of the region ${\cal V}_i(R)+{\cal V}_{t,i}(R)$
the functions ${\cal A}$ reduce to the breakup amplitude $A$,
since  in ${\cal V}_b(R)$ ${\cal F}^{(z)}(x_i,\hypfi_i)\ra 1$ ($z=c$, $n$).
The factor ${\cal F}^{(c)}$ takes into
account the effect of the Coulomb and centrifugal potential in
${\cal V}_{t,i}(R)$ and goes to zero in ${\cal V}_i(R)$.
Conversely the factor ${\cal F}^{(n)}$ takes into account the effects
of the NN potential and the centrifugal barrier. Since for $\ell_\b>0$
${\cal J}_{\ell_\b}$ vanishes in ${\cal V}_i(R)$, the influence of the
NN potential has been taken into account only in the $\ell_\b=0$ wave.

\section{The Kohn Variational Principle}
\label{sec:kvp}

The process we are interested in is a collision between a proton
and a deuteron, namely, the reaction
\be\label{eq:pd}
  (p+d)\rightarrow (p+d)+ (n+p+p)\ .
\ee
If the incident state of the $p-d$ system has
quantum numbers $L_0$, $S_0$ and $J$,  the elastic amplitude
in Eq.~(\ref{eq:pd4}) is characterized by
\be\label{eq:bez2}
 b_{LS}=\delta_{LL_0}\delta_{SS_0}\ .
\ee
The coefficient  $a_{LS}$ is usually renamed as  $-{\cal
S}_{L_0S_0,LS}$, ${\cal S}_{L_0S_0,LS}$ being the elements of the
elastic S--matrix. Since the breakup part should  not include
incoming waves, we must have
\be\label{eq:bez3}
  {\cal B}_{\a,T,\tauv}(x_i,\hypfi_i)=0\ \ .
\ee
The Eqs.~(\ref{eq:bez2}--\ref{eq:bez3}) give the boundary
conditions to be satisfied by the w.f. describing
the process~(\ref{eq:pd}). Explicitly, the asymptotic amplitude is
\bea
  \psi(i) &=& \sum_{LS} \Phi^d_{LS}(i)
       \biggl[ -{\cal S}_{L_0S_0,LS}
        {\cal H}^+_L(\eta_0,q_0\d_i)
       + \delta_{LL_0}\delta_{SS_0}  {\cal H}^-_L(\eta_0,q_0\d_i)
        \biggr] \nonumber \\
  && +   {  e^{\ii Q\rho -\ii \hat \chi \ln(2Q\rho) } \over \rho^{5\over 2}}
   \left (  \sum_{\a,T,\tauv}
    {\cal A}_{\a,T,\tauv}(\rho,\Omega_i)\sai\right )\ ,
    \quad \rho\ge R
   \ .\label{eq:pd5}
\eea

Let us now consider the integral
\be\label{eq:kohn1}
  I=\langle
  \Psi^{(-)}_{L_0S_0} |(H-E)|\overline{\Psi}_{L_0S_0}
   \rangle_R - \langle
  \overline{\Psi}^{(-)}_{L_0S_0}| (H-E)|\Psi_{L_0S_0}
  \rangle_R\ ,
\ee
where $\langle\rangle_R$ stands for the integration over
$d^3x_i\;d^3y_i$ of the
six--dimensional volume with $\rho\le R$ ($R\rightarrow\infty$)
and $\Psi^{(-)}$ is the time--reversed w.f. In the end the limit
$R\rightarrow\infty$ is taken.  The function  $\overline \Psi$ is a trial 
w.f. which will be chosen to satisfy
the boundary conditions. Without loss of generality, the trial w.f. can
be decomposed into three amplitudes, 
$\overline\Psi=\sum_{i=1}^3\overline\psi(i)$. The asymptotic
behavior of $\overline\psi(i)$ can be written  as
\bea
  \overline\psi(i) &=& \sum_{LS} \Phi^d_{LS}(i)
       \biggl[ -\overline {\cal S}_{L_0S_0,LS}
        {\cal H}^+_L(\eta_0,q_0\d_i)
       + \delta_{L_0,L} \delta_{S_0,S} 
         {\cal H}^-_L(\eta_0,q_0\d_i)
        \biggr]\nonumber \\
 && \qquad +
  {  e^{\ii Q\rho -\ii \overline \chi\log2 Q\rho} \over \rho^{5\over 2}}
   \left (  \sum_{\b,T,\tauv}  \overline {\cal A}_{\b,T,\tauv}(\rho,\Omega_i)
   \sai \right )\ ,\quad\rho\ge R\ .
   \label{eq:trial1}
\eea
The overlined quantities are free
parameters which represent approximations to the exact values.
It should be noticed that an {\it approximate expression}
$\overline \chi$ of the operator $\hat \chi$  has been introduced in the
above equation.

The important point is that the validity of the KVP for the elastic
part of the S--matrix can be proved also in this case. However, the
``choice'' of  $\overline\chi$ is not completely arbitrary, since
the matrix element $\langle\overline\Psi| (H-E)|\overline\Psi\rangle$
has to be finite. Moreover, in a practical calculation the convergence
of the elastic S--matrix should be checked whenever an improvement is
made to $\overline\chi$.

For example, in the case considered in Sect.~\ref{sec:phh} the
operator $\overline\chi$ is represented as an expansion over the
Hyperspherical Harmonic functions and therefore has not any
singular behavior for $\hypfi_i\ra\pi/2$, $i=1,2,3$. Moreover,
$\overline\chi\ra\hat \chi$ as the number of terms included
in the expansion basis is increased. Such a  procedure may result in a slow
convergence of $\overline{\cal A}$ towards the exact ${\cal A}$, in
particular for $\hypfi_i\ra\pi/2$, $i=1,2,3$. However, 
this problem has no practical consequences for the calculation of the
elastic S--matrix, as will be shown in Sect.~\ref{sec:phh}.

Contributions to the integral $I$ come only from the differential operators
present in $H$. By integrating by parts, $I$ takes contribution
only from the hypersurface $[\Omega]$ at $\rho=R$. Let us denote
with $[\Omega_x]$ the parts of $[\Omega]$ corresponding to a
particular asymptotic region ${\cal V}_x(R)$. Therefore,
\be\label{eq:kohn2}
 I =\sum_{i=1}^{3}  \left (I_i+I_{t,i}\right ) + I_b\ ,
\ee
where $I_i$, $I_{t,i}$ and  $I_b$ are the contributions coming from
$[\Omega_{i}]$,  $[\Omega_{t,i}]$ and $[\Omega_{b}]$, respectively.
On the hypersurface
$[\Omega]$, the trial and exact wave functions have reached the
asymptotic behavior given in Eqs.~(\ref{eq:pd5}) and~(\ref{eq:trial1}).
In $[\Omega_{i}]$, only the elastic part contributes, with the result
\be\label{eq:kohni5}
  I_i = {2\ii\over q_0}\k\htm \Bigl (\overline{\cal S}_{L_0S_0,L_0S_0}
         -  {\cal S}_{L_0S_0,L_0S_0}\Bigr) \ .
\ee
In fact, the breakup part goes like $\rho^{ -5/2}$ and 
gives a contribution $\propto R^{-3/2}$ which vanishes in the limit
$R\ra\infty$.
The contribution $I_{t,i}$ coming from the region $[\Omega_{t,i}]$ is
also zero. First of all, only breakup waves can contribute since
the elastic wave vanishes outside ${\cal V}_i(R)$. Moreover,
the ``dimension'' of the surface $[\Omega_{t,i}]$
is proportional to $R^{2+3\mu}$, whereas the integrand
is proportional to $1/R^5$. Thus, for $R\ra\infty$ the
integral in $[\Omega_{t,i}]$ behaves $\propto R^{3(\mu-1)}\ra0$. Finally,
in the case in which only outgoing breakup waves are present
in both the trial and exact w.f. the contributions of the two terms in
Eq.~(\ref{eq:kohn1}) cancel each other.

In $[\Omega_{b}]$,  $I_b$
reduces to
\be
  I_b = - \htm \int_{[\Omega_b]} d\Omega \biggl [
   \rho^{5\over 2} \Psi_{L_0S_0} {d \over d \rho}
    \rho^{5\over 2} \overline \Psi_{L_0S_0}
  -  \rho^{5\over 2} \overline \Psi_{L_0S_0}
    {d \over d \rho} \rho^{5\over 2} \Psi_{L_0S_0}
   \biggr]_{\rho=R}
  \ ,\label{eq:kohnb3}
\ee
and, using the asymptotic behaviors given in
Eqs.~(\ref{eq:pd4}) and~(\ref{eq:trial1}),
$I_b\propto 1/ R\ra 0$ for $R\ra\infty$.

Summing up all the contributions and remembering that
$(H-E)\Psi_{L_0,S_0}=0$ the {\it exact} expression is obtained:
\be\label{eq:kohn7}
   {\cal S}_{L_0S_0,L_0S_0} =
         \overline {\cal S}_{L_0S_0,L_0S_0} - {mq_0\over 6\ii\k\hbar^2}
    \langle \overline\Psi^{(-)}_{L_0S_0} |(H-E)|
    \overline \Psi_{L_0S_0}\rangle
    + {mq_0\over 6\ii\k\hbar^2}
    \langle \epsilon^{(-)}_{L_0S_0} |(H-E)|
    \epsilon_{L_0S_0}\rangle
   \ ,
\ee
where $\epsilon_{L_0S_0}=\overline\Psi_{L_0S_0}-\Psi_{L_0S_0}$ is the
``error'' introduced in the trial w.f. It can be noticed that
Eq.~(\ref{eq:kohn7}) is variational in
character, since the last term is quadratic in the error function
$\epsilon_{L_0S_0}$. Therefore, the functional
\begin{equation}\label{eq:kohnf}
  [ {\cal S}_{L_0S_0,L_0S_0}] =
         \overline {\cal S}_{L_0S_0,L_0S_0} - {mq\over 6\ii\k\hbar^2}
    \langle \overline\Psi^{(-)}_{L_0S_0} |(H-E)|
    \overline \Psi_{L_0S_0}\rangle
   \ ,
\end{equation}
represents a second order estimate for the exact diagonal element 
${\cal S}_{L_0S_0,L_0S_0}$.
The search of stationary solutions through the equation
\be\label{eq:kohn8}
  \delta [ {\cal S}_{L_0S_0,L_0S_0}] = 0   \ ,
\ee
provides the optimum  estimate. Here $\delta$ denotes the variation with 
respect to all the
trial parameters entering $\overline\Psi_{L_0S_0}$, including
$\overline S$ and $\overline {\cal A}$.
Moreover, the KVP can be readily
generalized to obtain variational  relations for the
non--diagonal matrix elements ${\cal S}_{L_1S_1,L_2S_2}$
of the elastic collisional matrix.

Clearly, the  principle is meaningful only if the two matrix elements
entering Eq.~(\ref{eq:kohn7}) are both {\it finite}. This is not
evident since for $\rho>R$  both $\overline \Psi$ and $\epsilon$ have
oscillatory behaviors.
However, since Eq.~(\ref{eq:kohn7}) is exact
and ${\cal S}$ and $\overline {\cal S}$ are finite, it is
sufficient to check  that one of the matrix elements is finite, in
particular $\langle\overline\Psi| (H-E)|\overline\Psi\rangle$. This imposes
some constraints on $\overline \chi$, which usually has
to be chosen coherently with the expansion basis used to construct
$\overline \Psi$ (see Sect.~\ref{sec:phh} for a practical example).

Applications of the KVP usually start with the expansion of the w.f. in
a certain basis which is truncated at some level.
The stability of the functional (\ref{eq:kohnf}) is studied in terms of
the dimension of the basis and in terms of the nonlinear parameters (if any).
Spurious resonances or spurious solutions in the convergence procedure could
appear due to a zero eigenvalue present in the operator $H-E$~\cite{schw61}.
These spurious solutions have been studied for a long time and it was
shown that the  KVP in its complex form, as presented here,
is well-behaved~\cite{schn88}.
Spurious solutions could appear in this case too, but these occur for
unrealistic
values of the nonlinear parameters, which are in general present in the
form of a regularizing factor or inside some exponent~\cite{lucc89}.
Below the DBT, a practical rule to verify the presence of
spuriousities is to apply the principle to the reactance matrix $K$
and its inverse $K^{-1}$ (inverse KVP) and verify the
relation $K\;K^{-1}=I$. 
Similarly, for energies above the DBT, one could apply the 
variation principle for the T-matrix (which can be derived by
following an analogous procedure to the one described here). 
With two independent estimates of (the elastic part of) the S-- and
T--matrices one can test, for example,
the relation $S=I-2\ii T$.  Besides these anomalies, which constitute 
an inherent problem of the method, the trial w.f. must be flexible enough to
approach the exact one in the asymptotic
and the internal regions. In other words, the error introduced due to
the approximate representation of the operator $\hat \chi$ in the truncated
basis should be reduced accordingly with the improvement in the
description of the internal part of the wave function.

The situation is rather similar to that which occurs when an
approximate form $\bar \phi_d$ of the deuteron w.f.
is used in Eq.~(\ref{eq:suri}). It is easy to
show that the KVP is still valid,
but the ``error'' term also includes
a quadratic term in $\bar\phi_d-\phi_d$~\cite{D72}.
In general, what happens is the following:
when the inaccuracies of $\bar \phi_d$ or $\overline\chi$
are ``small'',
the resulting errors in the elastic S--matrix elements  are of second
order, and can usually be disregarded. If
the inaccuracies are not small
enough, the expansion basis $\{{\cal Q}\}$
used to describe the internal
part of the w.f.  will spread out in the
asymptotic region in an attempt to remedy the defects of the
$\bar\phi_d$ and $\overline \chi$. If this happens,  the convergence
with respect to $\{{\cal Q}\}$ may be extremely slow, or it could even
happen that no convergence could be found.

\section{A Practical Example}
\label{sec:phh}

In this Section, the application of the pair--correlated
hyperspherical harmonic (PHH) technique to describe $p-d$ scattering 
for energies above the DBT is discussed.
In the PHH approach, the  trial w.f. is written as
\be\label{eq:chh1}
 \overline\Psi_{L_0S_0}= \Psi_A+ \Psi_C\ ,\qquad
 \Psi_A=\sum_{i=1}^3 \overline\psi_A(i)\ ,\qquad
 \Psi_C=\sum_{i=1}^3 \overline\psi_C(i)\ .
\ee
The first  term  $\Psi_A$ describes the system
when the two incident clusters are well separated and it is written in
the form  
\be\label{eq:chh3}
  \psi_A(i) = \sum_{LS} \Phi_{LS}(i)
       \biggl[ -{\cal S}_{L_0S_0,LS} \widetilde{\cal H}^+_L(\eta_0,q_0\d_i)
              + \delta_{LL_0}  \delta_{SS_0}
             \widetilde{\cal H}^-_L(\eta_0,q_0\d_i)
    \biggr] \ ,
\ee
where
\be\label{eq:coul2}
  \widetilde {\cal H}_L^\pm(\eta_0,q_0 \d_i)=
 { (1-e^{-\gamma \d_i})^{2L+1}\;
 G_L(\eta_0,q_0 \d_i)\pm
{\rm i} F_L(\eta_0,q_0 \d_i) \over q_0 \d_i}\ .
\ee
The factor $(1-e^{-\gamma d_i})^{2L+1}$
has been introduced to regularize the function $G$ at the origin,
and  $\gamma$ is taken as a nonlinear variational parameter.
Its choice is not a critical problem due to the large flexibility of the
trial w.f. in the internal region. The range of values
$\gamma=0.3-0.5$ fm$^{-1}$
has been found to be  adequate in all the cases studied.
For $d_i\gg 1/\gamma$, $\psi_A(i)$ coincides with the first term of
the asymptotic form~(\ref{eq:trial1}).

The second term $\Psi_C$ of the trial w.f. must describe those
configurations of the 
system where the particles are close to each other. For large
interparticle separations and energies below the
DBT, $\Psi_C$ goes to zero, whereas for higher energies it must
reproduce an outgoing three particle state. Each amplitude $\psi_C(i)$
is expanded in terms of the PHH
basis~\cite{KRV94,KRV97,KRV99},
\be\label{eq:chh2}
     \psi_C(i)=  \rho^{-5/2}\sum_{\alpha,T,\tauv} \sum_{K}
      u_{\a,T,\tauv,K}(\rho) f_\a(x_i)
      {}^{(2)}P^{\ell_\alpha,L_\alpha}_K(\hypfi_i)
      \sai\ ,
\ee
where ${}^{(2)}P^{\ell_\alpha,L_\alpha}_K$ is a
hyperspherical polynomial, defined for example in Ref.~\cite{F83},
$u_{\a,T,\tauv,K}(\rho)$ are the hyperradial functions to be determined
by the variational procedure and $f_\a(x_i)$ is a correlation function.
In the ``uncorrelated'' hyperspherical harmonic expansion (HH)
one has $f_\a(x_i)=1$, whereas
in the PHH approach it is included in order to better take into
account those correlations introduced by the repulsion of the
potential at short
distances. The correlation function is calculated by solving a two-body
Schroedinger--like equation and it goes to one for large values of
the interparticle distance $x_i$. In Eq.~(\ref{eq:chh2}), the sum over
$\a,T,\tauv$ is truncated after the inclusion of $N_c$ channels.
Usually all the channels with $\ell_\a+L_\a\le L_{\rm max}$, where 
$L_{\rm max}=6$ or $7$, are included in the expansion. For each
channel, the sum over the 
index $K$ of the hyperspherical polynomial runs from $K=0,1,2,\ldots$ 
up to a maximum value $K_{\a,T,\tauv}$. The latter values are
chosen in order to achieve the required convergence in the quantities
of interest. The total number of basis functions is therefore
\be\label{eq:MM}
  M= \sum_{\a,T,\tauv}^{N_c} (K_{\a,T,\tauv}+1)\ .
\ee

To simplify the notation let us label the basis
elements with the index $\mu\equiv[\a,T,\tauv,K]$,
and introduce the following completely antisymmetric correlated
spin-isospin-hyperspherical basis elements
\be\label{eq:bco}
     {\cal P}_\mu(\rho,\Omega)= \sum_{i=1}^3
      f_\a(x_i) {}^{(2)}P^{\ell_\alpha,L_\alpha}_K(\hypfi_i)
      \sai \ , 
\ee
which depends on $\rho$ through the correlation factor and form a
non--orthogonal basis.
In terms of $  {\cal P}_\mu(\rho,\Omega)$ the internal part is written as
\be
   \Psi_C= \rho^{-5/2}\sum_{\mu=1}^{M}
    u_{\mu}(\rho) {\cal P}_\mu(\rho,\Omega) \ .
\ee
The ``uncorrelated'' basis elements ${\cal P}^0_\mu(\Omega)$
are obtained from Eq.~(\ref{eq:bco}) in which all the correlation
functions are $f_\a(x_i)=1$. It is important to note that also 
the elements ${\cal P}^0_\mu(\Omega)$ do not form an orthogonal
basis, as has been discussed in Ref.~\cite{KMRV97}. In particular,  those
elements having the same grand--angular quantum number $G=\ell_\alpha +
L_\alpha + 2K$  but belonging to different channels are not
orthogonal among themselves. Moreover, some
${\cal P}^0_\mu(\Omega)$ with the same
$G,\Lambda_\alpha,\Sigma_\alpha,\tauv$ quantum numbers are
linearly dependent. In Ref.~\cite{KMRV97} these
states have been identified and eliminated from
the expansion.

In the present case, the basis elements
${\cal P}_\mu(\rho\ra\infty,\Omega)$ reduce to the
uncorrelated ones ${\cal P}^0_\mu(\Omega)$ in the
asymptotic region ${\cal V}_b(R)$. Therefore, it will be useful to
combine the correlated basis~(\ref{eq:bco}) in order to
define a new basis with the property of being orthonormal for $\rho=\infty$.
This can be readily accomplished by noting  that
\be
   n_{\mu\mu'}(\rho)= \int d\Omega\;
     {\cal P}_\mu(\rho,\Omega)^\dag
     {\cal P}_{\mu'} (\rho,\Omega) \ra N^{(0)}_{\mu\mu'}
    +{  N^{(3)}_{\mu\mu'}\over \rho^3}+{\cal O}(1/\rho^5)
      \ ,\qquad {\rm for\ }\rho\ra\infty\ ,  \label{eq:n}
\ee
where, in particular,
\be
   N^{(0)}_{\mu\mu'}= \int d\Omega\;
      {\cal P}^0_\mu (\Omega)^\dag
      {\cal P}^0_{\mu'} (\Omega)\ .  \label{eq:n1}
\ee
Let us define a matrix $U$ such that $ U \; N^{(0)} \; U^t=
{\cal D}$ is a diagonal matrix with diagonal elements ${\cal D}_\mu$
either $1$ or $0$. 
The ${\cal D}_\mu=0$ correspond to the states ${\cal P}^0_\mu
(\Omega)$ linearly dependent on the others.
New uncorrelated and correlated bases are defined as:

\be\label{eq:Qor}
   {\cal Q}^0_\mu(\Omega) \equiv \sum_{\mu'=1}^M U_{\mu\mu'} {\cal
   P}^0_\mu(\Omega)\ ,\qquad
   {\cal Q}_\mu(\rho,\Omega) \equiv \sum_{\mu'=1}^M U_{\mu\mu'} {\cal
   P}_\mu(\rho,\Omega)\ ,
\ee
The basis functions ${\cal Q}_\mu(\rho,\Omega)$ are
still not orthogonal for any finite values of $\rho$. 
When $\rho\ra\infty$, the elements 
${\cal Q}_\mu(\rho,\Omega)\ra {\cal Q}^0_\mu(\Omega)$. 
Due to the fact that some of the uncorrelated elements ${\cal
P}^0(\Omega)$ are linearly dependent, some elements
${\cal Q}^0_\mu (\Omega)$ are identically zero. Therefore, some
correlated elements
have the property: ${\cal Q}_\mu(\rho,\Omega) \ra 0$ as $\rho\ra\infty$.

In terms of the new basis, the internal part  $\Psi_C$ is simply
\be\label{eq:Qor2}
   \Psi_C=\rho^{-5/2}
          \sum_{\mu=1}^M \omega_\mu(\rho) {\cal Q}_\mu (\rho,\Omega)
   \ ,
\ee
where $\omega_\mu=\sum_{\mu'}(U^{-1})_{\mu'\mu}\; u_{\mu'}$.

The variation of the functional~(\ref{eq:kohnf}) with respect to the
hyperradial functions $\omega_\mu(\rho)$, which are the unknown quantities
entering in the description of the internal part of the w.f. $\Psi_C$,
leads to the set of inhomogeneous second order differential equations
given in Ref.~\cite{KRV97}, where the method of solution has been
discussed. For $\rho\ra\infty$, neglecting terms going to zero faster
than $\rho^{-1}$, the asymptotic expression of such a  set of differential
equations reduces to the form
\be \label{eq:c0}
   \sum_{\mu'} \biggl\{
   - \htm  \left( {d^2\over d\rho^2} +Q^2 \right ){\cal D}_\mu
    \delta_{\mu,\mu'} +
     {2\;Q\; c_{\mu,\mu'}\over \rho} \;
      +{\cal O}({1\over\rho^2})\biggr\}\omega_{\mu'}(\rho) = 0 \ ,
\ee
where the matrix $c$ is defined as
\be\label{eq:c}
   { c}_{\mu,\mu'}= \int d\Omega\;
      {\cal Q}^0_{\mu} (\Omega)^\dag
     \; \hat \chi \;
      {\cal Q}^0_{\mu'} (\Omega)
      \ .
\ee
In practice, the functions $\omega_\mu(\rho)$ are chosen to be
regular at the origin ($\omega_\mu(0)=0$) and, in accordance to
Eq.~(\ref{eq:c0}), to have the following behavior
\be\label{eq:asy2}
  \omega_\mu(\rho) \rightarrow
   \sum_{\mu'=1}^M
  \left ( e^{-\ii { c} \ln 2 Q\rho} \right)_{\mu,\mu'}\;
   \overline {\cal S}^{\;b}_{L_0S_0,\mu'} \; e^{\ii Q\rho} \ ,
   \qquad (\rho\ra\infty)\ ,
\ee
where $\overline {\cal S}^{\;b}_{L_0S_0,\mu'}$ are unknown
coefficients. Eq.~(\ref{eq:asy2}) reproduces the behavior of
the breakup term with
\be\label{eq:c2}
  \sum_{\a,T,\tauv}
    \sum_{i=1}^3 \overline {\cal A}_{\a,T,\tau}(\rho,\Omega_i)\sai
    =\sum_{\mu=1}^M  \overline {\cal S}^{\;b}_{L_0S_0,\mu}
     {\cal Q}_\mu (\rho,\Omega)\ , 
\ee
or 
\be\label{eq:c2b}
    \overline {\cal A}_{\a,T,\tau}(\rho,\Omega_i) =
      \sum_{K=0}^{K_{\a,T,\tauv}} \left(
      \sum_{\mu'=1}^M  \overline {\cal S}^b_{L_0,S_0,\mu'}
      U_{\mu'\mu} \right ) f_\a(x_i)
      {}^{(2)}P^{\ell_\alpha,L_\alpha}_K(\hypfi_i)
      \ ,\qquad \mu\equiv[\a,T,\tauv,K]\ ,
\ee
and
\be\label{eq:c3}
  \overline \chi = \sum_{\mu=1}^M \sum_{\mu'=1}^M
      {\cal Q}^0_{\mu} (\Omega)^\dag\;
     c_{\mu,\mu'}\;
      {\cal Q}^0_{\mu'} (\Omega)
    \ .
\ee
From the above equation it is clear that an enlargement of the PHH basis
by increasing $N_c$ and $K_{\a,T,\tauv}$ improves the
trial w.f. both in the internal and in the asymptotic regions.
The choice in Eqs.~(\ref{eq:asy2}) and~(\ref{eq:c3}) is natural
in the calculation scheme outlined here 
and ensures that the matrix element
$\langle \overline\Psi^{(-)}_{L_0S_0} |(H-E)|
 \overline \Psi_{L_0S_0}\rangle$ be finite . In fact
\be\label{eq:c4}
  <\overline\Psi^{(-)}_{L_0S_0}|H-E|\overline\Psi_{L_0S_0}>=
   \int_0^\infty d\rho\; {\cal I}(\rho)\ ,
\ee
and for  $\rho\ra\infty$ one has
\bea
  {\cal I}(\rho)&\ra& \htm
  \sum_{\mu=1}^M \sum_{\mu'=1}^M \biggl\{
  - \omega_\mu(\rho)\;
    \left( {d^2\over d\rho^2} +Q^2 \right ){\cal D}_\mu
    \delta_{\mu,\mu'} \omega_{\mu'}(\rho)+ \nonumber \\
  && +  \omega_{\mu}(\rho)\; {2Q c_{\mu,\mu'}\over \rho} \;
   \omega_{\mu'}(\rho) +{\cal O}({1\over\rho^2})\biggr\} \ra
    {\cal O}({1\over\rho^2}) \ .
    \label{eq:c5}
\eea

In conclusion, the procedure appears to be  well founded and in the
(hypothetical) limit of an infinite expansion it leads to the
exact results. The breakup amplitudes $\overline {\cal A}_\a(\rho,\Omega_i)$
are obtained in first order approximation in terms  of the PHH basis elements
${\cal Q}_\mu(\rho,\Omega)$, via Eq.~(\ref{eq:c2b}).
The convergence of $\overline {\cal A}(\rho,\Omega_i)$ as a
function of $N_c$ and $K_{\a,T,\tauv}$ 
has been found to be very slow when $\hypfi_i\ra\pi/2$. This is related
to the difficulties for the expansion~(\ref{eq:c2b}) of reproducing
the $x_i$ dependence of the  breakup amplitudes in the region ${\cal
V}_i(R)+{\cal V}_{t,i}(R)$ specified in
Eqs.~(\ref{eq:vb1}--\ref{eq:vb3}).  However, this 
problem has no consequences on the calculation of the elastic S--matrix
elements $[{\cal S}]$, since the regions $[\Omega_i]$ and
$[\Omega_{t,i}]$ do not give any contribution to the error term
$<\epsilon|H-E|\epsilon>$.

Once proved that the procedure can lead to the correct solution of the
problem, there are two main questions to be taken care of in practical
applications, namely:

\noindent 1. 
The boundary conditions given in Eq.~(\ref{eq:asy2}) are valid only at
$\rho=\infty$. However, such a problem can be overcome by using the
procedure proposed in Ref.~\cite{KRV97}. 
First of all, a set of $M$ coupled differential equations 
are derived from the variational condition $\delta_{\omega_\mu}[{\cal
S}]=0$. The functions $\omega_\mu(\rho)$ are determined in the region  
$\rho>\rho_0$ by expanding in powers of $1/\rho$ and verifying the
boundary conditions of Eq.~(\ref{eq:asy2}). Values of 
$\rho_0\approx 100$ fm have been found to be appropriate.
Then, the differential equations are solved numerically
in the region $(0,\rho_0)$ imposing the continuity of the logarithmic
derivative at $\rho=\rho_0$ with the solutions found for $\rho>\rho_0$.
This procedure fixes the values of the coefficients $\overline {\cal
S}^{\; b}_{L_0S_0,\mu}$. The stability of the 
solution has been tested by choosing different values of the matching
radius in the 
range $80 - 200$ fm. Finally, the variational condition
$\delta_{{\overline S}_{L_0S_0,LS}}[{\cal S}]=0$ is used to 
find the elastic coefficients $\overline {\cal S}_{L_0S_0,LS}$.

\noindent 2.
The convergence of the functional $[{\cal S}]$
with $N_c$ and $K_{\a,T,\tauv}$.
In all the calculations performed~\cite{KRV99},
the convergence pattern of $[{\cal S}]$
above the DBT was found to be rather similar to that
observed below the threshold~\cite{benchII}.
To give an idea of the convergence,  the calculated S--matrix
elements for $p-d$ scattering at $E_{lab}=(3/2)E=5$ MeV and $10$ MeV
are reported    in Table~\ref{tab:cons}. The state considered is
$J^\Pi=1/2^+$, which is in general the most structured one since
it must satisfy the orthogonality condition with respect 
to the three-nucleon bound state.
In this case the S--matrix is a $2\times2$ (complex) matrix
corresponding to the $p-d$ asymptotic states $L=0$, $S=1/2$ and $L=2$,
$S=3/2$. The NN interaction considered is the AV18
potential~\cite{AV18}.

In Table~\ref{tab:cons}, the real and complex part
of the two phase--shifts $^2S_{1/2}$, $^4D_{1/2}$ and the mixing
parameter $\eta_{1/2+}$ are given for different values of the number
of channels $N_c=8,18,26$. In the first case ($N_c=8$) only  the 8
channels given in Table~\ref{tab:chan3}  have been
considered
since they give the major contribution to the bound state.
The cases $18$ and $26$ correspond to including
all the $\tauv=1/2$ channels with $\ell_\a+L_\a \leq4$ and $6$,
respectively. In all the cases, the number of hyperspherical states
for each channel has been increased until the convergence was reached.
It has been found that a number of states similar to the one used 
for energies below the DBT is sufficient to get converged
results, i.e.
$8$ basis elements for the first $4$ channels, $6$ basis elements for
channels $5-14$ and $4$ basis elements for the successive channels.

Also the convergence with $N_c$ is analogous to that
obtained for energies below the deuteron breakup, as can be seen
in Table~\ref{tab:cons}. The first
$8$ channels give the most important contribution to
the S--matrix, with results accurate up to three digits.
Higher order channels give minor contributions, of the order of a few
percent. The contribution of the channels with $\tauv=3/2$ has been
found to be very small and it has been disregarded in the calculation
presented here.

\section{Conclusions}
\label{sec:conc}

The purpose of this paper is to demonstrate the validity of
the  KVP to describe $p-d$ elastic scattering above the
DBT. The practical application of the principle has been
analyzed, with particular reference to the use of the PHH expansion
technique.
It was shown that the principle remains formally unchanged when the breakup
channels are open. In this case, the asymptotic behavior of the
w.f. has been extensively discussed with special care in
the treatment of the long range Coulomb potential. Different asymptotic 
regions has been introduced for different values of the hyperspherical 
variables ($\rho,\theta_i$) and the solution of the Faddeev-Noble
equations has been studied in each case. 
It was found that certain
asymptotic configurations in the $p+p+n$ system have negligible
probability. For example, the breakup w.f. vanishes in region
${\cal V}_i$ when particles $j,k$ are protons or when they are
in a relative state with angular momentum $\ell_\alpha>0$, independently of
the fact that they form a $p-p$ or an $n-p$ pair. 
The transition from the solution in region ${\cal V}_b$ to these
vanishing solutions has been studied.

Moreover, it was explicitly shown that
the use of ``approximate'' terms in the asymptotic logarithmic distortion 
of the breakup wave function does not influence the formal derivation and 
the practical
application of the principle. This is due to the fact that 
$\exp(-\ii \hat \chi \ln(2Q\rho))\times  {\cal A}_{\a,T,\tauv}$ is in
any region a smooth function (as discussed in Sect.~\ref{sec:summ}).
Thus it is not very critical to try to reproduce it with some
approximate method like an expansion over HH functions.

The hyperspherical basis has been found to be a rather natural one
to describe  
the asymptotic behavior of the wave function describing scattering
above the DBT, including also the long--range Coulomb interaction.
In fact, the equations to be solved have a
very simple structure in terms of $\rho$ in the asymptotic region
and, accordingly, analytical solutions can be obtained. In practical
applications it is not difficult to reduce the problem
to the determination of the hyperradial functions in
the range $0\le \rho \le \rho_0$ with the application of specific
boundary conditions at $\rho_0$. Appropriate values of the matching radius
$\rho_0$ have been found to be in the range of $80\div 200$ fm.

The convergence of the quantities of interest is the usual
problem which one has to face when applying a variational
procedure. In two preceding papers~\cite{KRV97,KRV99}, it has been
shown that the convergence of the elastic observables can be achieved
with the same precision as below DBT. Realistic NN and 3N interactions
can be used to calculate the observables in the $n-d$ and $p-d$ case, so
meaningful comparisons with experimental data can be
performed.  In the present work
a detailed discussion of the problem related to 
possible inaccuracies introduced from an approximate description
of the outgoing breakup waves was presented, 
including a numerical example.

In conclusion, the KVP provide a formalism from which it is possible
to obtain reliable and fully--converged results for the elastic part
of the $p-d$ S--matrix at energies above the DBT even with the inclusion of
the long range Coulomb interaction. Applications of the method 
described here were already reported in Refs.~\cite{KRV99,kie99} where
the elastic 
differential cross section and the vector and tensor analyzing powers
were studied at $E_{lab}=5$ MeV and $10$ MeV.
The further problem related to a precise determination of
the breakup amplitudes ${\cal A}_{\a,T,\tauv}$ is currently under
progress and it
will be discussed elsewhere~\cite{KRV99b}.

\appendix
\section*{}
\label{sec:appa}

In this appendix, the asymptotic behavior of the (outgoing) breakup wave
of $\psi(i)$ in ${\cal V}_i(R)$ is given. As discussed in
Sect.~\ref{sec:Vip},  only the $\ell_\a=0$ channels of the components
$\psi_j(i)$ and $\psi_k(i)$ do not vanish in ${\cal
V}_i(R)$ and we have
\bea  
   \psi_i(i) &=& 0 \ , \nonumber \\
  \psi_j(i) & = & C(\rho) \sum_{\Lambda S\Sigma}
    (-)^S { f_{\Lambda S \Sigma}(x_i) \over \sqrt{2}\; x_i } 
     \widetilde {\cal Y}_{0\Lambda\Lambda S\Sigma}(i) \ ,
     \qquad {\rm region}\ {\cal V}_i(R)\ , 
     \label{eq:ax1}  \\
  \psi_k(i) & = & C(\rho) \sum_{\Lambda S\Sigma}
    { g_{\Lambda S \Sigma}(x_i) \over \sqrt{2}\; x_i } 
     \widetilde {\cal Y}_{0\Lambda\Lambda S\Sigma}(i) \ ,
     \nonumber
\eea
where $C(\rho)$  and $\widetilde {\cal
Y}_{\ell,L,\Lambda,S,\Sigma}(i)$ are defined in 
Eqs.~(\ref{eq:ee}) and~(\ref{eq:ang2}), respectively. The sum over the
channels has been restricted to those with $\ell=0$ as stated
previously. The factors $(-)^S$ and $1/\sqrt{2}$ have been introduced
for convenience. The functions $f$ and $g$ are determined as explained in
the following. First of all, $ g_{\Lambda S \Sigma}= 
f_{\Lambda S \Sigma}$ since $\psi(i)$ must be antisymmetric under the 
exchange $j\leftrightarrow k$. Therefore, the amplitude 
$\psi(i)= \psi_i(i)\; \Xi_i  +  \psi_j(i)\; \Xi_j  + \psi_k(i)\;
\Xi_k$ can be cast in the form
\be
   \psi(i)=C(\rho) \sum_{\Lambda S\Sigma}
    {  f_{\Lambda S \Sigma}(x_i) \over x_i } 
    \overline {\cal Y}_{\Lambda S \Sigma}(i) 
    \ ,     \label{eq:ax2}  
\ee
where 
\be\label{eq:ax3} 
     \overline {\cal Y}_{\Lambda S \Sigma}(i) =
     \widetilde {\cal Y}_{0\Lambda\Lambda S\Sigma}(i) 
     |(t^j t^k)_{T,0} (t^i)_{{1\over 2},+{1\over 2}}\ket
    \ ,\quad T=1-S\ . 
\ee
The amplitude $\psi(i)$ given in Eq.~(\ref{eq:ax2}) is easily seen to be
antisymmetric  with respect to the exchange $j\leftrightarrow k$. 

The amplitudes $\psi(j)$ and $\psi(k)$ in the
region ${\cal V}_i(R)$ can be obtained in the same way as in
Sect.~\ref{sec:Vj}, where the behavior of $\psi(i)$ in ${\cal
V}_j(R)$ and ${\cal V}_k(R)$ is discussed. In this region the following
relations hold
\be\label{eq:app5a}
 \x_j\rightarrow -\k \y_i\, \quad \y_j\rightarrow - {1\over 2} \y_i\
 ,\quad \hypfi_j\rightarrow{\pi\over 6}\ ,
\ee
and
\be\label{eq:app5b}
 \x_k\rightarrow  \k \y_i\, \quad \y_k\rightarrow - {1\over 2} \y_i
 \ ,\quad \hypfi_k\rightarrow{\pi\over 6}\ .
\ee
Accordingly, $\psi_{j\ {\rm or}\ k}(j)$ and $\psi_{j\ {\rm or}\ k}(k)$
satisfy the FNE with $V^{\rm NN}=0$ and coincide with the ``free''
solution~(\ref{eq:pd4c}). On the other hand, as discussed in
Sect.~\ref{sec:Vj}, the components $\psi_i(j)$ 
and $\psi_i(k)$ vanish in ${\cal V}_i(R)$ due to the Coulomb
repulsion. Therefore, in region ${\cal V}_i(R)$
\bea
  \psi(j) &=& C(\rho)  
   \sum_{\a} \widetilde A^j_{\a,j}({\pi\over 6}) \sbj\;\Xi_j
            +\widetilde A^j_{\a,k}({\pi\over 6}) \sbj\;\Xi_k \ ,
             \label{eq:app6a} \\
  \psi(k) &=& C(\rho)  
   \sum_{\a} \widetilde A^k_{\a,j}({\pi\over 6}) \sbk\;\Xi_j
            +\widetilde A^k_{\a,k}({\pi\over 6}) \sbk\;\Xi_k \ ,
             \label{eq:app6b}
\eea
where $\a=\{\ell,L,\Lambda,S,\Sigma\}$.

Since, under the exchange $j\leftrightarrow k$, it must be verified
that $\psi(j)\ra -\psi(k)$ and vice versa, the amplitudes   
$\widetilde A^{j\ {\rm or}\ k}_{\a,{j\ {\rm or}\ k}}(\pi/6)$ are not
independent.  Using the previous relations for the Jacobi coordinates,
then
\bea
 \sbj&\ra& \sum_{S'} (-)^{S'} D_{\a,S'} \widetilde {\cal
    Y}_{0\Lambda\Lambda S'\Sigma}(i)\ ,\label{eq:app7a} \\
 \sbk&\ra& \sum_{S'} (-)^{\ell+S} D_{\a,S'} \widetilde {\cal
    Y}_{0\Lambda\Lambda S'\Sigma}(i)\ ,\label{eq:app7b}
\eea
where
\be\label{eq:app8}
 D_{\a,S'}= (-)^{S+S'} \hat \ell \hat L \hat S \widehat {S'}
        \left( \begin{array}{ccc}
        \ell & L & \Lambda \\
        0 & 0 & 0 
       \end{array}   \right)     
      \left\{ \begin{array}{ccc}
        {1\over 2} & {1\over2} & S \\
        \Sigma & {1\over2}  & S' 
       \end{array}   \right\} \ , 
\ee
and $\hat \ell=\sqrt{2\ell+1}$, etc.
Note that only states with orbital angular momentum $\ell=0$ enter
the r.h.s of Eqs.~(\ref{eq:app7a}) and~(\ref{eq:app7b}).
Finally, we can write
\be\label{eq:app8b}
 \psi(j)+\psi(k) = 
   C(\rho) \sum_{\Lambda S' \Sigma} {\cal K}_{\Lambda S'\Sigma}\;
      \overline {\cal Y}_{\Lambda S'  \Sigma}(i) \ ,
      \qquad {\rm region}\ {\cal V}_i(R)\ , 
\ee
where $\overline {\cal Y}_{\Lambda S'  \Sigma}(i)$ is given in
Eq.~(\ref{eq:ax3})  and
\be\label{eq:app9}
 {\cal K}_{\Lambda S'\Sigma} = \sum_{\ell L S}
     \Bigl(\widetilde A^j_{\a,j}({\pi\over 6})+(-)^{S'}
           \widetilde A^j_{\a,k}({\pi\over 6}) \Bigr)
          D_{\a,S'} \sqrt{2} \ .
\ee
Inserting the expressions~(\ref{eq:ax2})
and~(\ref{eq:app8b}) in the FNE and disregarding all terms ${\cal
O}(1/y_i)$ or  ${\cal O}(1/\rho)$, the following equation is obtained
\be\label{eq:app11}
  -\htm  f^{\prime\prime}_{\Lambda S \Sigma}(x_i) +
   V_{2S+1}(x_i) f_{\Lambda S \Sigma}(x_i) = 
   -x_i V_{2S+1}(x_i)\; {\cal K}_{\Lambda S  \Sigma}\ ,
\ee
where
\be\label{eq:app10}
  \int d\hat x_i\; d\hat y_i\;
  \Bigl( \overline {\cal Y}_{\Lambda' S' \Sigma'}(i)\Bigr)^\dag
   V^{\rm NN}(\x_i) 
  \overline {\cal Y}_{\Lambda S  \Sigma}(i) 
 = V_{2S+1}(x_i) \delta_{SS'}\delta_{\Lambda\Lambda'}
  \delta_{\Sigma\Sigma'}\ .
\ee
$V_{2S+1}(x_i)$ is the projection of the NN potential acting in
a state with orbital angular momentum zero, spin $S$ and isospin $T=1-S$.

The general solution of Eq.~(\ref{eq:app11}) can be written as the sum of
the solution of the homogeneous part and a particular solution
of the complete equation. The solution of the homogeneous
part will be denoted as $\gamma_{\Lambda S \Sigma}
f^h_{2S+1}(x_1)$, where $\gamma_{\Lambda S \Sigma}$ 
is an arbitrary coefficient; $f^h_{2S+1}$ coincides with the (regular)
solution of the zero--energy Schroedinger equation of a
two--nucleon system in the $2S+1$ s-wave state. For large $x_i$ values
it goes like $x_i-a_{2S+1}$, where $a_{1}$ ($a_3$) is the singlet
(triplet) scattering length.

A particular solution is simply
\be\label{eq:app12}
    \tilde f^p_{\Lambda S \Sigma}= - x_i\; {\cal K}_{\Lambda S \Sigma}
     \ . 
\ee
Note that the particular solution exactly cancels 
the amplitudes $\psi(j)+\psi(k)$ in ${\cal V}_i(R)$. In fact,
\bea\label{eq:app13}
  \Psi &=& \psi(i)+\psi(j)+\psi(k) \ , \nonumber \\
       &\approx& C(\rho) \sum_{\Lambda S\Sigma} \gamma_{\Lambda S \Sigma}
          { \tilde f^h_{\Lambda S \Sigma}(x_i) \over x_i } 
          \overline {\cal Y}_{\Lambda S \Sigma}(i) \ .
\eea
Namely, in ${\cal V}_i(R)$ the pair $jk$ is in a relative s--wave
state with vanishing relative velocity, whereas the motion of the
third particle $i$ is governed by the distorted wave $C(\rho)\approx
C(y_i)$.

For large values of $x_i$, $f_{\Lambda S \Sigma}(x_i) /x_i \ra {\rm const}$. 
In the region ${\cal V}_{t,i}(R)$, the
amplitude given in Eq.~(\ref{eq:ax1}) should smoothly match the
behavior of $\psi_j(i)$ and $\psi_k(i)$. Therefore,  the constants
$\gamma_{\Lambda S \Sigma}$ are related to the breakup amplitudes
$\widetilde A^i_{\a,j}$ via the relations
\be\label{eq:app13b}
 {(-)^S\over\sqrt{2} } (\gamma_{\Lambda S \Sigma} - 
  {\cal K}_{\Lambda S \Sigma} ) = 
  \widetilde A^i_{0\Lambda\Lambda S\Sigma,j}(\pi/2)  
  \ .
\ee
Finally, the functions $\phi^0_\a(x_i)$ introduced in
Eq.~(\ref{eq:pd11}) have been defined to be
\be\label{eq:app14}
  \phi^0_{0\Lambda\Lambda S\Sigma}(x_i)= 
      {(-)^S\over\sqrt{2} } 
    {1\over  \widetilde A^i_{{0\Lambda\Lambda S\Sigma},j}(\pi/2) }
     { f_{\Lambda S \Sigma}(x_i) \over x_i}
     \ .
\ee

\begin{table}[p]
\label{tab:cons}
\begin{tabular}{cccc}
$E_{lab}=5$ MeV & $N_c=8$& $N_c=18$     & $N_c=26$ \\
\hline
$^4D_{1/2}$  &(-5.45,0.004)&(-5.43,0.004)&  (-5.43,0.004) \\
$^2S_{1/2}$  &(-42.0,1.74) &(-41.8,1.74) &  (-41.8,1.74)  \\
$\eta_{1/2+}$&(1.06,-0.03) &(1.06,-0.03) &  (1.05,-0.03) \\
\hline
$E_{lab}=10$ MeV & $N_c=8$& $N_c=18$     & $N_c=26$ \\
\hline
$^4D_{1/2}$   &(-7.41,0.23)&(-7.31,0.24) & (-7.30,0.24) \\
$^2S_{1/2}$   &(-60.8,11.7)&(-60.6,11.7) & (-60.6,11.7) \\
$\eta_{1/2+}$ &(1.04,0.06) &(1.02, 0.06) & (1.01,0.06)
\end{tabular}
\caption{Convergence of the eigenphase shift and mixing angle
parameters of elastic $p-d$ scattering  as a function of the number
of channels included in the PHH expansion of the internal part of the
w.f. The state considered has quantum number $J^\Pi=1/2^+$. The
nuclear potential model considered is the AV18 interaction.
$E_{lab}$ is the laboratory energy of the incident proton (the center
of mass energy is $E=2/3 E_{lab}$).}
\end{table}

{\arraycolsep 8pt
\begin{table}
\begin{tabular}{cccccccc}
\hline
    $\alpha$ & $\ell_{\alpha}$ & $L_{\alpha}$  & $\Lambda_\alpha$ &
      $S_{\alpha}$ &  $T$ & $\Sigma_\alpha$ & $\tauv$
        \\
\hline
  1  & 0 & 0 & 0 &   1 & 0 & {1/ 2} &  1/2     \\
  2  & 0 & 0 & 0 &   0 & 1 & {1/ 2} &  1/2      \\
  3  & 0 & 2 & 2 &   1 & 0 & {3/ 2} &  1/2      \\
  4  & 2 & 0 & 2 &   1 & 0 & {3/ 2} &  1/2      \\
  5  & 2 & 2 & 0 &   1 & 0 & {1/ 2} &  1/2      \\
  6  & 2 & 2 & 2 &   1 & 0 & {3/ 2} &  1/2      \\
  7  & 2 & 2 & 1 &   1 & 0 & {1/ 2} &  1/2      \\
  8  & 2 & 2 & 1 &   1 & 0 & {3/ 2} &  1/2      \\
\hline
\end{tabular}
\caption[Table]{\label{tab:chan3}
Quantum numbers for the first $8$
channels considered in the expansion of the internal part  of the
three--nucleon scattering w.f. }
\end{table}
}

\end{document}